\documentclass[10pt,journal]{IEEEtran}

\usepackage{amsmath}
\usepackage{amsthm}
\usepackage{amssymb}
\usepackage{mathrsfs}
\usepackage{mathtools}
\usepackage{bm}
\usepackage{subfigure}   
\usepackage{indentfirst} 
\usepackage{cite} 
\usepackage{graphicx}
\usepackage{epstopdf}
\usepackage{acronym}
\usepackage[bookmarks,colorlinks]{hyperref}
\usepackage{comment}
\usepackage{url}

\newcommand{\Verify}[1]{\footnote{\textcolor{red}{#1}}}
\newcommand{\Revise}[1]{\textcolor{red}{#1}}

\newtheorem{mypro}{Proposition}

\newtheorem{lemma}{Lemma}

\acrodef{far}[FAR]{frequency agile radar}
\acrodef{hrr}[HRR]{high range resolution}
\acrodef{cs}[CS]{compressed sensing}
\acrodef{crr}[CRR]{coarse range resolution}
\acrodef{sar}[SAR]{synthetic aperture radar}
\acrodef{isar}[ISAR]{inverse \ac{sar}}

\title{Phase Transitions in Frequency Agile Radar Using Compressed Sensing}
\author{Yuhan Li, Tianyao Huang$^*$, Xingyu Xu, Yimin Liu, and Yonina C. Eldar
	\thanks{Part of this paper was presented in the IEEE radar conference 2020 \cite{Li2020}. This work was supported by the National Natural Science Foundation of China under Grants 61801258. 
		 T. Huang, Y. Liu, Y. Li and X. Xu are with the EE Department, Tsinghua University, Beijing, China. 
		Y. C. Eldar is with the Faculty of Math and CS, Weizmann Institute of Science, Rehovot, Israel. 
		$^*$Correspondence: huangtianyao@tsinghua.edu.cn.
}}

\begin{document}
	\maketitle
	
	\begin{abstract}
		\Ac{far} has improved anti-jamming performance over traditional pulse-Doppler radars under complex electromagnetic circumstances. To reconstruct the range-Doppler information in FAR, many compressed sensing (CS) methods including standard and block sparse recovery have been applied. 
		In this paper, we study phase transitions of range-Doppler recovery in FAR using CS. In particular, we derive closed-form phase transition curves associated with block sparse recovery and complex Gaussian matrices, based on prior results of standard sparse recovery under real Gaussian matrices. 
		We further approximate the obtained curves with elementary functions of radar and target parameters, facilitating practical applications of these curves. Our results indicate that block sparse recovery outperforms the standard counterpart when targets occupy more than one range cell, which are often referred to as extended targets. 
		Simulations validate the availability of these curves and their approximations in FAR, 
		which benefit the design of the radar parameters. 
	\end{abstract}
	
	\acresetall

	\section{Introduction}
	\indent \Ac{far} varies its carrier frequencies randomly in a pulse by pulse manner. It synthesizes a wide bandwidth by coherently processing echoes of different frequencies, achieving \ac{hrr} while requiring only a narrow-band hence low-cost receiver \cite{axelsson2007analysis}. This facilitates applications including \ac{sar} \cite{yang2012random} and \ac{isar} imaging \cite{wang2019phase,LiuZhenAES2014}. In addition, 
	 \ac{far} possesses excellent electronic counter-countermeasures performance \cite{axelsson2007analysis}, supports spectrum sharing \cite{Ma2020joint}, and enhances spectrum efficiency \cite{cohen2017spectrum}. Owing to these advantages, \ac{far} has drawn considerable attention in the radar community \cite{CaoAES2020,HuangMajorcom2020}.
	
	\ac{far} relies on signal processing algorithms to recover the range-Doppler parameters of observed targets and clutter. Early works \cite{axelsson2007analysis} employed the traditional matched filtering for range-Doppler reconstruction, which led to significant sidelobe pedestal. As a consequence, weak targets could be covered by the sidelobe of dominant ones or strong clutter \cite{liu2008range}. To alleviate the sidelobe pedestal problem, \ac{cs} methods (also known as sparse recovery \cite{eldar2012compressed}) have been suggested, which exploit the inherent sparsity of the targets \cite{liu2008range,huang2018analysis}. 
	The authors in \cite{wang2019theoretical} further extended the standard sparse recovery approach to block sparse recovery to account for the situation of extended targets, where a target may be larger than the range resolution and therefore can occupy more than one range cell \cite{gerlach1999adaptive}.
	In this situation, the scatters of a target share the same Doppler effect and gather along range, leading to block sparsity\cite{eldar2015sampling}. When confronted with such block sparse situations including extended targets, block sparse recovery is considered to perform better than the conventional sparse recovery\cite{eldar2010block,mishali2008reduce}.
	
	Precise conditions that guarantee reconstruction in FAR have been considered in several papers. 
	In \cite{huang2018analysis} and \cite{wang2019theoretical}, the authors provided sufficient conditions (in terms of the numbers of targets $ K $, radar pulses $ N $ and available frequencies $ M $) that guarantee successful reconstruction of target scenes with high probability using standard and block sparse recovery, respectively. 
	Nevertheless, those conditions, based on the well-known coherence property, are generally loose and pessimistic \cite{carin2011coherence}, and are therefore not accurate enough to predict the actual recovery performance given the radar and target parameters.
	
	To obtain a tighter bound, we study here the phase transition, 
	which emerges in many convex optimization problems \cite{amelunxen2014living}. In the content of \ac{cs}, phase transition means that there exist thresholds that divide the plane of parameters, i.e., the number of observations and the sparsity level, into regions, where recovery succeeds and fails with high probability \cite{foucart2013sparse}. These thresholds are called phase transition curves.
	Finding analytical expression of these curves is an active area. 
	For standard sparse recovery, bounds on the phase transition curve of $\ell_1$ norm minimization under standard Gaussian matrices were established in \cite{donoho2006high,stojnic2009various}. Generalization to block sparse recovery and for complex number form were given in \cite{4813252} and \cite{yang2011phase}, respectively. However, these approximate results assume the observation matrices to be large, and have complicated form, resulting in difficulty to apply in practice. A more concise and tight bound, which has no requirement on the size of the observation matrix, was given in \cite{amelunxen2014living} using integral geometry. Nevertheless, it is confined to standard sparse recovery under real-valued Gaussian matrices.

	In FAR, block sparse recovery is preferred, and measurement matrices are complex-valued. Therefore, we first extend the results of \cite{amelunxen2014living} to block sparse situations and complex Gaussian matrices. 
	While existing analyses are based on the Gaussian assumption, measurement matrices in \ac{far} are not Gaussian but structured. Empirical experiments show that many random matrices exhibit identical phase transition curves as Gaussian matrices\cite{donoho2009observed}. 
	We demonstrate numerically in Section \ref{section:sim} that the obtained bounds derived from Gaussian matrices are tight and accurate for \ac{far}. Thus our results provide more precise conditions for exact range-Doppler reconstruction, compared to former works \cite{huang2018analysis,wang2019theoretical}. 
	
	Next, we approximate the obtained bounds, which involve minimization over an integral function, with some elementary functions. In particular, under relatively sparse scenes where there are only a few extend targets, we show that the required numbers of measurements when using the block and standard sparse recovery are on the order of $ 2MK + O\left( K\sqrt{M \log \frac{N}{K\sqrt{M}}}\right)  $ and $ 2MK + O\left( KM \log\frac{N}{K}\right)  $, respectively. The former requires less radar measurements for exact reconstruction of extended targets, since $ \sqrt{M \log \frac{N}{K\sqrt{M}}} < M \log\frac{N}{K}  $ for reasonably large $N$ and $M$. 
	The accuracy of these approximations is validated by simulations. These approximations not only simplify the calculation of the bounds, facilitating their use in practical scenarios, but also explicitly and quantitatively reveal the dependency of the required number of measurements on the radar and target parameters. These explicit results enable theoretical performance comparison between the block and standard sparse recovery methods, which demonstrate the superiority of block sparse recovery.
	
	We summarize the main contributions of this paper as follows.
	\begin{itemize}
		\item  We derive the phase transition curves under Gaussian matrices for block sparse recovery, which are also numerically accurate for \ac{far}, associated with structured and non-Gaussian matrices. Therefore, we provide more precise conditions on exact range-Doppler reconstruction than previous works. 
		\item  We approximate the phase transition curves of both block and standard sparse recovery with some elementary functions, facilitating their use in practical \ac{far}, and demonstrating that block sparse recovery outperforms the standard one when reconstructing extended targets.
	\end{itemize}
	
	The rest of the paper is structured as follows. Section~\ref{section:FAR} introduces the signal model of FAR. In Section~\ref{section:pre}, we briefly review basic concepts of standard and block sparse recovery as well as phase transitions for standard sparse recovery. Section~\ref{section:PT in BSR} extends the results of phase transitions in Section~\ref{section:pre} to block sparse recovery and complex problems and Section~\ref{section:Ph in FAR} analyzes phase transitions for the FAR model, which are verified by simulations in Section~\ref{section:sim}. Section~\ref{section:conclusion} concludes the paper.
	
	Throughout the paper, we use $\mathbb{R}$ and $\mathbb{C}$ to denote the real and complex number set, respectively. For $x \in \mathbb{R}$, $\lfloor x \rfloor$, represents the largest integer no greater than $x$. Vectors are written as lowercase boldface letters (${\rm e.g.,}\ \bm{a}$), while matrices are written as uppercase boldface letters (${\rm e.g.,}\ \bm{A}$). For a vector $\bm{a}$, $\|\bm{a}\|_{i}$ denotes the $\ell_i$ norm of $\bm{a}$ and $j\coloneqq \sqrt{-1}$. Given a matrix $\bm{A}$, $[\bm{A}]_{m,n}$ is the $(m,n)$-th entry of $\bm{A}$. The transpose operator is $(\cdot)^T$ and $\mathbb{E}(\cdot)$ means the expectation of a random value. The real and imaginary parts of a complex-valued argument are written as $\Re \cdot$ and $\Im \cdot$, respectively. We use $\mathscr{N}(0,1)$ to denote standard Gaussian distribution.\vspace{-3ex}

	\section{FAR Signal Model}\label{section:FAR}
	We begin by reviewing the signal model of FAR in Subsection \ref{subsection:Radarmodel}, followed by its matrix form in Subsection \ref{subsection:Radarmatrix}.
	
	\subsection{Radar Model}\label{subsection:Radarmodel}
	We introduce the signal model of FAR following \cite{huang2018analysis} and \cite{wang2019theoretical}. We start with the expressions of the transmissions and the received echoes from a single scattering point, representing target or clutter. We then extend the echo model to the case of multiple targets/clutter.
	
	A FAR system changes the frequencies from pulse to pulse. Suppose that the radar transmits $N$ monotone pulses during a coherent processing interval (CPI). 
	The carrier frequency of the $n$-th pulse can be written as $f_n = f_c + C_n \Delta f$, where $f_c$ represents the initial frequency, $\Delta f$ represents the frequency step, and $C_n$ is the $n$-th random modulation code. We assume that all the modulation codes are independently and identically distributed (i.i.d.) random variables with uniform density on $\mathcal{M}:=\{0,1,\ldots,M-1\}$, i.e., $C_n \sim U(\mathcal{M})$. 
	Then the $n$-th transmitted pulse, $n \in \mathcal{N}:=\{ 0,1,\ldots,N-1\}$, is written as\vspace{-1ex}
	\begin{equation}\label{eq:transmit}
		S_{T} (n,t) = {\rm rect}\left((t-nT_r)/{T_p}\right)e^{j2\pi f_{n} (t-nT_r)},\vspace{-1ex}
	\end{equation}
	where $T_r$ is the pulse repetition interval (PRI), $T_p$ is the pulse duration, and ${\rm rect}(t)$ equals 1 when $0\le t\le 1$ and 0 otherwise. 

	The received echoes can be seen as delays of transmissions. To clearly present the signal model, we assume that there exists only one ideal scatterer with complex scattering coefficient $\zeta$, which has an initial range $r$ from the radar and is moving along the radar's line of sight at a fixed speed of $v$. The time delay between the received and the transmitted signal at time instant $t$ takes the form $\tau (t) \coloneqq \frac{2(r+vt)}{c} $, where $c$ is the velocity of light. Under the stop-and-hop assumption \cite{huang2020multi}, it holds that $\tau(t) \approx \tau(nT_r)$ during the $n$-th pulse. Thus, the received echo of the $n$-th transmission can be expressed as\vspace{-1ex}
	\begin{equation}\label{eq:recive}
		S_{R}(n,t) = \zeta S_T(n,t-\tau (t))\approx\zeta S_T(n,t-\tau (nT_r)).\vspace{-1ex}
	\end{equation}
	
	After down-conversion, the received echoes become\vspace{-1ex}
	\begin{small}
	\begin{equation}\label{eq:downconversion}
		S_{D}(n,t) = S_R(n,t)e^{-j2\pi f_n (t-nT_r)}.\vspace{-1ex}
	\end{equation}
	Substituting \eqref{eq:recive} and \eqref{eq:transmit} into \eqref{eq:downconversion} and rearranging, we obtain \vspace{-1ex}
	\begin{equation}\label{eq:downconversion2}
		\begin{aligned}
			&S_{D}(n,t) =\zeta {\rm rect}\left(\frac{t-nT_{r}-\tau (nT_r)}{T_p}\right)e^{-j2\pi f_{n}\tau(nT_r)}\\
			&=\ddot{\zeta}{\rm rect}\left(\frac{t-nT_{r}-\tau (nT_r)}{T_p}\right)e^{j2\pi f_{r}C_{n}+j2\pi f_v \epsilon_n n},\vspace{-2ex}
		\end{aligned}
	\end{equation}
	\end{small}
	where 
	$\epsilon _n \coloneqq 1+\frac{C_n \Delta f}{f_c}$. The parameters $\ddot{\zeta}:=\zeta e^{-j4\pi f_{c} r/c}$, $f_r \coloneqq -\frac{2\Delta fr}{c}$ and $f_v \coloneqq -\frac{2f_c vT_r}{c}$, representing the effective intensity, range frequency and velocity frequency, respectively, are unknown and need to be estimated.
	
	The down-converted echoes are sampled at the Nyquist rate of a single pulse, $f_s = \frac{1}{T_p}$, at time instances, $t = nT_r +\frac{l_s}{f_s}$, $l_s = 0,1,\ldots,\left\lfloor T_r f_s\right\rfloor$. Each sample corresponds to a \ac{crr} bin. The coarse range will be refined by estimating $f_r$ or $r$ from the echoes. Here, $f_r$ and $r$ are referred as the \ac{hrr} information.
	Since data from those bins are processed identically and individually, without loss of generality, we assume that the target is located in the $l_s$-th CRR bin, and that the target does not move outside the bin during the CPI. Consequently, the ${\rm rect}(\cdot)$ term in \eqref{eq:downconversion2} at the $l_s$ sampling instance equals 1, so that sampled $S_D(n,t)$ becomes \vspace{-1ex}
	\begin{equation}\label{equa:single}
		S_{S}(n)\coloneqq S_D \left(n,nT_r +l_s/f_s\right)=\ddot{\zeta}e^{j2\pi f_r C_n +j2\pi f_v \epsilon_n n}.\vspace{-1.2ex}
	\end{equation}

	The model \eqref{equa:single}, derived for the case of a single scatterer, can be extended to the setting in which $K$ targets/clutter exist in a CRR bin. Particularly, we denote by $v_k$ and $f_{v_k}$ the velocity and velocity frequency of the $k$-th target (or clutter unit), respectively, with $k = 0,1,\ldots,K-1$, and assume that the $k$-th target is composed of $Q_k$ scatterers, moving at the same speed while the ranges and scattering intensities are different. For the $i$-th scatterer of the $k$-th target, ${\ddot{\zeta}}_{ki}$, $r_{ki}$ and $f_{r_{ki}}$ denote the scattering coefficient, initial range and range frequency, respectively. We then extend \eqref{equa:single} into a model for multiple targets as follows,\vspace{-1.5ex}
	\begin{equation}\label{equa:return model}
		S_{S}(n) = \sum_{k=0}^{K-1}\sum_{i=0}^{Q_k -1}\ddot{\zeta}_{ki}e^{j2\pi f_{r_{ki}}C_{n}+j2\pi f_{v_k} \epsilon_n n}.\vspace{-1.5ex}
	\end{equation} 
	Here, $\{\ddot{\zeta}_{ki}, f_{r_{ki}}, f_{v_k}\}$ are unknown and need to be estimated from $S_{S}(n)$.
	
	In the next subsection, we arrange the signal model in matrix form, which suggests a block sparse recovery approach for range-Doppler reconstruction.\vspace{-2.5ex}

	\subsection {Signal Model in Matrix Form} \label{subsection:Radarmatrix}
	To write \eqref{equa:return model} in matrix form, we first divide the continuous $f_r$ and $f_v$, representing range and Doppler parameters into grid points. In particular, since $(f_r, f_v)$ is unambiguous in the domain $[0,1)^2$ and their resolutions are $1/M$ and $1/N$, respectively, we discretize $f_r$ and $f_v$ at the rates of $1/M$ and $1/N$, respectively, resulting in a series of grid points $({p}/{M} ,{q}/{N})$, $p\in \mathcal{M}$, $q \in \mathcal{N}$.
	
	We consider a discrete model, which assumes that all scatterers are situated exactly on the grid points, and use the matrix $\bm{X} \in \mathbb{C}^{M \times N}$ to encapsulate the scattering intensities, given by\vspace{-1.2ex}
	\begin{equation}
		[\bm{X}]_{p,q} \coloneqq \begin{cases}
			\ddot{\zeta}_{ki}, \quad \exists (k,i),\ {\rm s.t.} \ (f_{r_{ki}},f_{v_k})=(\frac{p}{M},\frac{q}{N}),\\
			0, \quad {\rm otherwise}.\vspace{-1.2ex}
		\end{cases}
	\end{equation}
	In practical scenes, scatterers of targets and clutter may be continuously distributed rather than located on the discrete grid. It is shown in \cite{wang2019theoretical} that radar returns of the discrete model well approximate the counterparts scattered from continuously located scatterers.  
	Let $\bm{x}_q \in \mathbb{C}^{M}$ denote the $q$-th column of $\bm{X}$, representing the \ac{hrr} profile of the target with Doppler frequency $q/N$, $q\in \mathcal{N}$. 
	Vectorizing $\bm{X}$ yields $\bm{x}\coloneqq [\bm{x}_0^T,\bm{x}_1^T,\ldots,\bm{x}_{N-1}^T]\in \mathbb{C}^{MN}$. Thus, $\bm{x}$ contains $N$ blocks each having $M$ entries, which represent the \ac{hrr} profile corresponding to a unique Doppler frequency \cite{wang2019theoretical}, and is called as having a \emph{block} structure. For the scenario that includes $K$ targets, at most $K$ blocks in $\bm x$ are nonzero, and $\bm x$ is a so-called $K$ block sparse vector. 
	
	Following \cite{wang2019theoretical}, we now write \eqref{equa:return model} in matrix form as \vspace{-1.5ex}
	\begin{equation} \label{equa:matrix form}
		\bm{y} = \bm{\Theta x},\vspace{-1.5ex}
	\end{equation}
	where the $n$-th entry of the measurement vector $\bm{y} \in \mathbb{C}^N$ is given by $[\bm{y}]_{n} := S_{S}(n)$. 
	The observation matrix $\bm{\Theta} \in \mathbb{C}^{N \times MN}$, in accordance with $\bm{x}$, is separated into $N$ blocks as 
	\vspace{-2.5ex}
	\begin{equation}\label{equa:FARblock}
		\bm{\Theta} \coloneqq [\bm{\Theta}_0,\bm{\Theta}_1,\ldots,\bm{\Theta}_{N-1}],\vspace{-1.5ex}
	\end{equation}
	where each block $\bm{\Theta}_q \in \mathbb{C}^{N \times M}$ has $(n,p)$-th entry given by\vspace{-1.5ex}
	\begin{equation}\label{equa:FARmodel}
		[\bm{\Theta}_q]_{n,p} = e^{j2\pi\frac{p}{M}C_n +j2\pi\frac{q}{N}\epsilon_{n}n},\quad q,n \in \mathcal{N}, p \in \mathcal{M}.\vspace{-1.5ex}
	\end{equation}
	
	
	In some scenarios, only a random selection out of all $ N $ pulses are transmitted (in the aim of, e.g., reducing power consumption), or partial observations in $ \bm y $ are abandoned and not processed because they have a strong interferer. This will lead to a compressive observation model, where only randomly selected entries of $ \bm y $ and $ \bm \omega $, as well as the corresponding rows of $ \bm \Psi $, remain.
		
	The matrix $\bm{\Theta}$ has more columns than rows, hence recovering $\bm{x}$ from $\bm{y}$ is an under-determined problem. Generally, there are only a small amount of targets occurring in a certain CRR, which means that $K$ is small, causing $\bm{x}$ to be block sparse \cite{eldar2010block}. 
	Therefore, block sparse recovery can be utilized to recover $\bm{x}$ and henceforth the target parameters. 
	We will review some basic concepts on sparse recovery in Section~\ref{section:pre}, and analyze the recovery performance in Section~\ref{section:PT in BSR}. 
	
	Since the purpose of this paper is to identify the fundamental limits on the recovery performance of FAR using sparse recovery methods, 
    throughout the paper we focus on the noise free model \eqref{equa:matrix form}, following typical approaches \cite{donoho2006high,stojnic2009various,amelunxen2014living,foucart2013sparse,4813252,yang2011phase}.\vspace{-2ex}


	\section{Preliminaries on Compressed Sensing}
	\label{section:pre}
	
	In this section, we briefly introduce some preliminaries on \ac{cs} and its phase transitions.
	In Section \ref{subsection:SR}, \ac{cs} methods including non-block and block sparse recovery are reviewed. Then in Section \ref{subsection:PT in SR}, we introduce the phase transition phenomenon in sparse recovery, which serves as a theoretical tool for performance analysis. 
	To distinguish between the specific radar parameters like $N$, we use lower case letters such as $n$ to denote the dimensions of matrices associated with a more general sparse recovery problem. Since some variables in this section can be either real or complex valued, and will be specified in later discussions, we use $\mathbb{S}$ to represent $\mathbb{R}$ or $\mathbb{C}$ for convenience. \vspace{-3ex}
	\subsection{Sparse Recovery}\label{subsection:SR}
	
	Consider an under-determined problem $\bm{y} =\bm{\Psi x}$, 
	where $\bm y \in \mathbb{S}^n$, $\bm \Psi \in \mathbb{S}^{n \times d}$ and $\bm x \in \mathbb{S}^d$. Here, $\bm x$ has no more than $s$ nonzero entries, and is called an $s$ sparse vector. \ac{cs} recovers $\bm x$ by harnessing the sparsity as in
	the following optimization program:
	\vspace{-1.5ex}
	\begin{equation}\label{eq:l0}
		\bm{\hat{x}}=\mathop{\arg\min}_{\bm{x}}\|\bm{x}\|_0, \text{ s.t. }\bm{y} =\bm{\Psi x}.\vspace{-1.5ex}
	\end{equation}
	Since the `$\ell_0$ norm' optimization \eqref{eq:l0} is NP-hard\cite{eldar2012compressed}
	, under appropriate conditions, the problem can be solved more efficiently by $\ell_{1}$ norm minimization, i.e., \vspace{-1.5ex}
	\begin{equation}\label{equa:l1}
		\bm{\hat{x}}=\mathop{\arg\min}_{\bm{x}}\|\bm{x}\|_1, \text{ s.t. }\bm{y} =\bm{\Psi x}.\vspace{-1.5ex}
	\end{equation}
	
	The standard sparse recovery problem \eqref{equa:l1} can be extended to block sparse recovery. With some abuse of notation, we consider a block-structured vector $\bm x \in \mathbb{S}^{md}$ consisting of $d$ blocks where each block has $m$ entries, denoted by $\bm x := \left[\bm x_0^T,\bm x_1^T,\dots,\bm x_{d-1}^T \right]^T$. Here, $\bm x_q \in \mathbb{S}^m$ denotes the $q$-th block. We use $s_B$ to represent the block sparsity, i.e., at most $s_B$ blocks in $\bm x$ are nonzero. 
	Similarly, we redefine $\bm \Psi \in \mathbb{S}^{n \times md}$ as the measurement matrix. To solve for $\bm x$ from observations $\bm y = \bm \Psi \bm x$, we exploit the block sparsity in $\bm x$ by considering an $\ell_{2,1}$ minimization problem \cite{eldar2010block} \vspace{-1.5ex}
	\begin{equation}\label{equa:l21}
		\bm{\hat{x}}=\mathop{\arg\min}_{\bm{x}}\|\bm{x}\|_{2,1}, \text{ s.t. }\bm{y} =\bm{\Psi x}.\vspace{-1.5ex}
	\end{equation}
	Here, the $\ell_{2,1}$ norm, given by $\|\bm{x}\|_{2,1}\coloneqq \sum_{q=0}^{d-1}\|\bm{x}_{q}\|_{2}$, is defined with respect to the block width $m$. When $m = 1$, \eqref{equa:l21} reduces to \eqref{equa:l1}.
	
	Both standard and block sparse recovery, i.e., \eqref{equa:l1} and \eqref{equa:l21}, can be applied to FAR, and provide unique solutions with high probability under certain conditions \cite{huang2018analysis, wang2019theoretical}. 
	Particularly, the recoverable number of targets $K$ is on the order of $O(\sqrt{\frac{N}{{\rm log}(MN)}})$ using $\ell_1$ norm minimization \cite{huang2018analysis} and $O(\frac{N}{M{\rm log}(MN)})$ using $\ell_{2,1}$ norm minimization\cite{wang2019theoretical}, where we recall that $N$ and $M$ are the numbers of pulses and available frequencies, respectively. These 
	obtained conditions are sufficient yet too pessimistic, as discussed in Section \ref{section:sim}.
	Consequently, these conditions are difficult to use directly in practical FAR systems to guide the waveform design under given parameters.
	To seek appropriate conditions that guarantee exact recovery with high probability and are tight enough to provide design criterion for radar systems, we resort to phase transition curves, which we introduce next.\vspace{-2ex}

	\subsection{Phase Transition Phenomenon}\label{subsection:PT in SR}
	In this section, we briefly introduce the phase transition phenomenon, following the ideas in \cite{amelunxen2014living}.
	
\begin{figure}[ht] 
		\centerline{\includegraphics[width=0.5\linewidth]{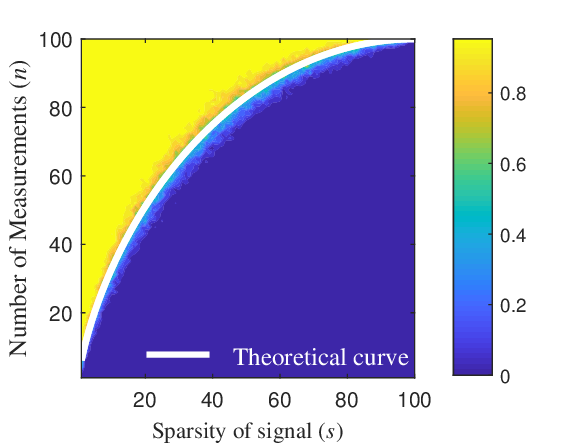}}\vspace{-1.5ex}
		\caption{Probabilities that \eqref{equa:l1} exactly solves $\bm x$ simulated from 50 trials. Here, $d = 100$, $\bm{\Psi}\in \mathbb{R}^{n \times d}$ has entries obeying i.i.d. $\mathscr{N}(0,1)$, and the nonzero entries of $\bm{x}$ are randomly 1 or -1. Exact recovery is proclaimed when the estimate $\hat{\bm{x}}$ satisfies $\|\hat{\bm{x}}-\bm{x}\|_2\le10^{-5}$. The theoretical curve is calculated analytically and is introduced in Proposition \ref{pro:phasel1}.}\vspace{-2ex}
		\label{fig1}
	\end{figure}

	Many CS works have focused on when exact recovery is possible using \eqref{equa:l1} and how these conditions change as a function of the problem parameters.
	Early works such as \cite{donoho2006high} observed that the probability of exact recovery possesses a phase transition with respect to the number of measurements, $n$, and the sparsity of the signal, $s$. Here, a phase transition means a dramatic change in the probability of exact recovery when these parameters, $n$ and $s$, vary around certain values. To illustrate the phase transition phenomenon associated with the probability of exact recovery, we empirically present in Fig.~\ref{fig1} the probabilities of exact recovery, assuming an under-determined real-valued Gaussian measurement matrix $\bm{\Psi}$, applying \eqref{equa:l1} under different pairs $(n,s)$. 
	The corresponding phase transition points $(n,s)$ where phase transitions happen, compose what we call the phase transition curve.
	This curve precisely characterizes the required conditions for exact recovery.
	While empirically calculating the curve is usually time consuming, expressions of the theoretical curve have been given in certain cases as we discuss next.\vspace{-3ex}

	The paper \cite{amelunxen2014living} identified a theoretical phase transition curve for a more general optimization problem of the form 
	\begin{equation}\label{equa:convex}
		\bm{\hat{x}}=\mathop{\arg\min}_{\bm{x}}f(\bm{x}), \text{ s.t. }\bm{y} =\bm{\Psi x},\vspace{-1.5ex}
	\end{equation}
	under a real-valued Gaussian measurement matrix $\bm \Psi$, based on integral geometry techniques. 
	Here, $f(\cdot)$ is restricted to be convex and does not take the value $-\infty$. In general, $f(\cdot)$ indicates the `structure' in a vector. For example, $f(\cdot) = \lVert \cdot \rVert_1$ characterizes the standard sparsity of a vector, and in this case \eqref{equa:convex} reduces to \eqref{equa:l1}. 
	To calculate the number of measurements $n$ which causes a phase transition, two concepts are introduced, the descent cone and the statistical dimension. The descent cone of a proper convex function $f:\mathbb{R}^d\rightarrow\mathbb{R}$ at the point $\bm{x} \in \mathbb{R}^d$ is defined as 
	\begin{equation}
		D(f,\bm{x})\coloneqq \mathop{\bigcup}_{\tau > 0}\{ \bm{y}\in \mathbb{R}^d: f(\bm{x}+\tau\bm{y}) \le f(\bm{x})\}.
	\end{equation}
	It depicts the conic hull of the perturbations which decrease or maintain $f$ around $\bm{x}$. 
	The number of measurements $n$ which causes a phase transition, representing the phase transition curve, depends on the descent cone at the point $\bm{x}$, given by
	\begin{equation}
		n = \delta(D(f,\bm{x})) = \mathbb{E}\left[ {\rm dist}^2(\bm{g},D(f,\bm{x}))\right].
	\end{equation}
	Here, $\delta(\cdot)$ is called the statistical dimension of  a cone, the expectation is taken over the random vector $\bm{g} \in \mathbb{R}^{d}$, obeying the Gaussian distribution $ \bm{g} \thicksim \mathscr{N}(\bm{0},\bm{I})$,  and ${\rm dist}(x,S)$ denotes the Euclidean distance from a vector $\bm{x} \in \mathbb{R}^d$ to a set $S\subset \mathbb{R}^d$
	\begin{equation}
		{\rm dist}(\bm{x},S)\coloneqq \inf\{\|\bm{x}-\bm{y}\|_{2}:\bm{y}\in S\}.
	\end{equation}
	Now, identifying the phase transition curve becomes calculating the statistical dimension with respect to the $\ell_1$ norm. 
	
	Directly calculating the statistical dimension is difficult, a tight upper bound on it is used instead in \cite{amelunxen2014living}. To obtain the bound, we first introduce the following definition. For an appropriate convex function ${f}: \mathbb{R}^d\rightarrow\mathbb{R}$, e.g., $\ell_{1}$ norm of a $d$ dimensional vector, the subdifferential $\partial {f}(\bm{x})$ at a point $\bm{x} \in\mathbb{R}^d$ is defined as
	\begin{equation}
		\partial {f}(\bm{x})\coloneqq \{{\bm s\in\mathbb{R}^d:f(\bm{y})\ge f(\bm{x})+\langle\bm{s},\bm{y}-\bm{x}\rangle, \ \forall \  \bm{y}\in\mathbb{R}^d\}},
		\label{eq:subdiff}
	\end{equation}
	\noindent where $\langle \cdot,\cdot\rangle$ denotes the dot product between two vectors. 
	Next, the following upper bound is derived in Proposition 4.1 of \cite{amelunxen2014living}: 
	
	\begin{mypro}[\hspace{1sp}\cite{amelunxen2014living}]
		\label{thm:upperbound}
		For $\bm{x} \in \mathbb{R}^{d}$ and a convex function ${f}: \mathbb{R}^d\rightarrow\mathbb{R}$, demanding the subdifferential $\partial{f}(\bm{x})$ to be compact, nonnull, and not containing the origin, the following function\vspace{-1ex}
		\begin{equation}\label{equa:J}
			J(\bm{x},f)\coloneqq \mathop{\inf}_{\tau\ge0}\mathbb{E}[{\rm dist}^{2}(\bm{g},\tau\cdot\partial {f}(\bm{x}))]\vspace{-1ex}
		\end{equation}
		upper bounds $\delta(D(f,\bm{x}))$, 
		where the expectation in \eqref{equa:J} is taken over the random vector $\bm{g} \in \mathbb{R}^{d}$, obeying the Gaussian distribution $ \bm{g} \thicksim \mathscr{N}(\bm{0},\bm{I})$.\vspace{-1ex}
	\end{mypro}

	To apply Proposition \ref{thm:upperbound}, we substitute $f(\cdot)$ with $\|\cdot\|_1$ in \eqref{equa:J} and calculate the infimum distance expectation $J(\bm{x},\|\cdot\|_1)$, which implies the upper bound on $\delta(D(\|\cdot\|_1,\bm{x}))$. We then express the upper bound as a function of the sparsity and the dimension of $\bm{x}$, $s$ and $d$, as claimed in the following proposition: \vspace{-1ex}
	
	\begin{mypro}[\hspace{1sp}\cite{amelunxen2014living}]
		\label{pro:phasel1}
		Given an $s$ sparse signal $\bm{x}\in\mathbb{R}^{d}$, $\varphi(s,d)$, defined as \vspace{-1.5ex}
		\begin{small}
			\begin{equation}
				\label{equa:sparse}
				\varphi(s,d) \!\!:=\!\! \mathop{\inf}_{\tau\ge0}\left\{s(1+\tau^{2})+(d-s)\int_{\tau}^\infty(u-\tau)^2\phi(u)\mathrm{d}u\right\},\vspace{-2ex}
			\end{equation}
		\end{small}
		upper bounds $\delta(D(\|\cdot\|_{1},\bm{x}))$, where $\phi(u)\coloneqq\sqrt{\frac{2}{\pi}}\exp(-\frac{u^2}{2}),u\ge0$, is the probability density function of the folded normal distribution.\vspace{-1ex}
	\end{mypro}
	\noindent Proposition \ref{pro:phasel1} offers a way to calculate the location of the phase transitions without performing complicated simulations running $\ell_1$ norm minimization algorithms. The upper bound $\varphi(s,d)$ in \eqref{equa:sparse} is tight, as shown in Fig.~\ref{fig1}, denoted by `theoretical curve'.\vspace{-2ex}

	\section{Phase Transitions in Block sparse recovery} \label{section:PT in BSR}
	Here we extend the result of \cite{amelunxen2014living} to the block sparse setting. 
	We present the theoretical results for real-valued Gaussian matrices in Subsection~\ref{subsection: PT in CNF}, and complex Gaussian matrices in Subsection~\ref{subsection: PT in CNF}.\vspace{-2ex}
	
	\subsection{Real-valued Cases}\label{subsection:PTC in BSR}
	Consider the model in \eqref{equa:l21} with $\bm y \in \mathbb{R}^n$, $\bm{\Psi} \in \mathbb{R}^{n\times md}$ and $\bm x \in \mathbb{R}^{md}$. Here, we assume that $\bm{\Psi}$ is a real-valued Gaussian matrix and $\bm x$ is block sparse with sparsity $s_B$. 
	According to Proposition \ref{thm:upperbound}, we let $f(\bm{x})$ be the $\ell_{2,1}$ norm of $\bm{x}$ with a block size $m$. Thus, the phase transition curve becomes $\delta(D(\|\cdot\|_{2,1},\bm{x}))$ and can be achieved by calculating the term in the right hand side of \eqref{equa:J} with respect to $f(\bm{x})$. 
	Following the steps used for the derivation of Proposition~\ref{pro:phasel1}, we obtain the following proposition, which provides an upper bound on $\delta(D(\|\cdot\|_{2,1},\bm{x}))$ in terms of $s_B$ and $d$.\vspace{-1ex}
	
	\begin{mypro}\label{pro:ph in bsr}
		Given an $s_B$ block sparse signal $\bm{x}\in\mathbb{R}^{md}$ having a block size $m$, the function $\varphi_m(s_B,d)$, defined as \vspace{-1.5ex}
		\begin{small}
			\begin{equation}\label{eq:block curve}
				\begin{aligned}
					&\varphi_m(s_B,d) \coloneqq\\ &\quad \mathop{\inf}_{\tau\ge0}\left\{s_B(m+\tau^{2})+(d-s_B)\int_{\tau}^\infty(u-\tau)^2\phi_m(u)\mathrm{d}u\right\},\vspace{-1ex}
				\end{aligned}
			\end{equation}
		\end{small}upper bounds $\delta(D(\|\cdot\|_{2,1},\bm{x}))$. Here, $\phi_m(u)$ is the probability density function of the $\chi$-distribution with $m$ degrees of freedom, given by\vspace{-1.8ex}
		\begin{equation}\label{equa:blocksparse}
			\phi_m(u) \coloneqq \begin{cases}\frac{u^{m-1}e^{-u^{2}/2}}{2^{m/2-1}\Gamma(\frac{m}{2})},\quad u\ge 0,\\
				0,\quad {\rm otherwise},\vspace{-1.5ex}
				
			\end{cases}
		\end{equation}
		where $\Gamma(\cdot)$ is the gamma function.\vspace{-1.5ex}
	\end{mypro}
	\begin{proof}
		See \ref{proof:bsr}.
	\end{proof}

	Proposition \ref{pro:ph in bsr} offers a theoretical bound on the phase transition curve in block sparse recovery, which is empirically tight as will be shown in Section \ref{section:sim} by experiments. It generalizes Proposition~\ref{pro:phasel1}, because standard $\ell_1$ norm minimization can be regarded as a special case of block sparse recovery with $m=1$ and $s_B = s$. In this case, the curve $\varphi_m(s_B,d)$ becomes identical to $\varphi(s,d)$ in Proposition~\ref{pro:phasel1}.
	Proposition~\ref{pro:ph in bsr} also paves the way to discussing sparse recovery with complex-valued measurement matrices.
	This is because both $\ell_1$ and $\ell_{2,1}$ norm minimization under complex-valued measurement matrices can be expressed by $\ell_{2,1}$ norm minimization in real-valued formulations, as discussed in the next subsection.\vspace{-2ex}

	\subsection{Complex-valued Cases}\label{subsection: PT in CNF}
	
	It is well known that complex-valued sparse recovery problems can be reformulated into real-valued ones \cite{yang2011phase}. 
	We use the same model \eqref{equa:l21} with Subsection~\ref{subsection:PTC in BSR} except that variables are complex valued: $\bm{y} \in \mathbb{C}^n$, $\bm{x}\in \mathbb{C}^{md}$, and $\bm{\Psi} \in \mathbb{C}^{n \times md}$ being a complex-valued Gaussian matrix. The model can be converted into a real-valued form $\bm{y}_r = \bm{\Psi}_r\bm{x}_r$ by introducing notations: $\bm{y}_r := \left[ \Re \bm{y}^T, \Im \bm{y}^T \right]^T \in \mathbb{R}^{2n}$, $\bm{\Psi}_r := \begin{bmatrix} \Re \bm{\Psi}&-\Im \bm{\Psi}\\ \Im \bm{\Psi}&\Re \bm{\Psi} \end{bmatrix} \in \mathbb{R}^{2n \times 2md}$ and $\bm{x}_r := \left[ \Re \bm{x}^T, \Im \bm{x}^T \right]^T \in \mathbb{R}^{2md}$. The $q$-th block in $\bm x$ is rewritten as $\bar {\bm{x}}_q = \left[ \Re \bm{x}_q^T, \Im \bm{x}_q^T \right]^T \in \mathbb{R}^{2m}$. We then exchange the entries in $\bm x_r$ such that we obtain $\bar {\bm{x}}: = \left[\bar {\bm{x}}_0^T, \dots, \bar {\bm{x}}_{d-1}^T \right]^T \in \mathbb{R}^{2md}$. Applying the same arrangement to the columns of $\bm \Psi_r$ yields $\overline{\bm \Psi} \in \mathbb{R}^{2n\times 2md}$, and we have $\bm{y}_r = \overline{\bm \Psi} \bar{\bm x}$. Here, the real-valued vector $\bar {\bm{x}}$ has $d$ blocks, which each contains $2m$ entries. The block sparsity of $\bar {\bm{x}}$ remains unchanged, $s_B$. 
	Define the $\ell_{2,1}$ norm of $\bar {\bm{x}}$ with respect to the block size $2m$, i.e., $\lVert \bar {\bm{x}} \rVert_{2,1} := \sum_{q=0}^{d-1} \lVert \bar{\bm{x}}_q \rVert_{2}$. Then, it can be verified that $\lVert {\bar {\bm{x}}}_q \rVert_{2} = \lVert  {\bm{x}}_q \rVert_{2}$ and hence $\lVert \bar {\bm{x}} \rVert_{2,1} = \lVert  {\bm{x}} \rVert_{2,1}$, where we recall that the latter $\ell_{2,1}$ norm is defined with respect to the block size of $m$. Consequently, the original complex-valued model \eqref{equa:l21} is equivalent to the optimization problem \vspace{-1.5ex}
	\begin{equation}
		\label{eq:realvalued}
		\hat{\bar{\bm x}} = \mathop{\arg\min}_{\bar{\bm{x}}}\|\bar{\bm{x}}\|_{2,1}, \text{ s.t. } \bm{y}_r = \overline{\bm \Psi} \bar{\bm 
			x}.\vspace{-1.5ex}
	\end{equation}
	Since $\ell_1$-norm based sparse recovery is a special case of block sparse recovery, following the same steps, we find that complex-valued $\ell_1$ norm minimization is equivalent to a real-valued $\ell_{2,1}$ norm minimization with block width of 2.

	We now calculate the bound on $\delta(D(\|\cdot\|_{2,1},\bm{x}))$ for the complex-valued case, based on the real-valued representation \eqref{eq:realvalued}. Proposition~\ref{thm:upperbound} assumes that entries in the measurement matrix are mutually independent Gaussian variables, while $\bm{\Psi}_r$ or $\overline{\bm \Psi}$ has duplicate entries, which are not independent. However, the dependence introduced by $\bm{\Psi}_r$ has little impact on its phase transition curve according to the empirical results in \cite{yang2011phase}, which inspires us to apply Proposition~\ref{pro:ph in bsr}, derived from Proposition~\ref{thm:upperbound}, for the real-valued optimization problem \eqref{eq:realvalued}. 
	Therefore, 
	the phase transitions of the optimization problem \eqref{eq:realvalued} emerge when $2n = \varphi_{2m}(s_B,d)$, where $\varphi_m(s,d)$ is defined in \eqref{eq:block curve}. We then denote by 
	\vspace{-2ex}
	\begin{equation}
		\label{eq:complexcurve}
		\varphi^{c}_m(s_B,d) \coloneqq \varphi_{2m}(s_B,d)/2
		\vspace{-1.5ex}
	\end{equation}
	an approximate bound on the phase transition curve. 
	The accuracy of \eqref{eq:complexcurve} will be verified in Section \ref{section:sim}. It provides the location of phase transitions in complex-valued block sparse recovery, which can be applied to standard sparse recovery as well.\vspace{-1ex}

	\section{Phase transition in FAR}\label{section:Ph in FAR}
	
	In this section, we adapt the derived phase transition curves to \ac{far}. We first present phase transition curves of \ac{far} using block sparse recovery, followed by the counterpart using standard sparse recovery. We approximate and simplify the expressions of these curves under certain assumptions, which facilitates the calculation of these curves. In Subsection~\ref{subsection:far block} and \ref{subsection:far standard}, we show the results of block and standard sparse recovery, respectively. A discussion of these results is presented in Subsection~\ref{subsection:far comp}.

	
	Phase transition curves of \ac{far} are inspired by Propositions~\ref{pro:phasel1}, \ref{pro:ph in bsr} and \eqref{eq:complexcurve}, which however are given under Gaussian matrices. 
	Generally, these curves are not theoretically applicable to FAR \eqref{equa:FARmodel}, because the measurement matrix in \eqref{equa:FARmodel} is not Gaussian but highly structured. Currently, there is no theoretical evidence that Proposition~\ref{thm:upperbound} holds for such structured measurement matrices.
	However, we will show in the next section by simulations that \eqref{eq:complexcurve} accurately indicates phase transitions in FAR.
	
	Consider a \ac{far} radar with the number of pulses and available frequencies being $N$ and $M$, respectively. The radar illuminates $K$ targets/clutter, of which each occupies all the $M$ \ac{hrr} bins and is regarded as a block of size $M$. These results can be simply extended to the special case when some targets only occupy partial \ac{hrr} bins.\vspace{-2ex}
	
	
	\subsection{Block Sparse Recovery}
	\label{subsection:far block}

	Assuming only $N_b$ observations out of the whole $N$ radar echoes are available, we use \eqref{eq:complexcurve} to identify the required $N_b$ for exact target reconstruction with block sparse recovery. Substituting $n = N_b$, $m = M$, $d = N$ and $s_B = K$ into \eqref{eq:complexcurve}, we have \vspace{-2.5ex}
	
	\begin{small}
		\begin{equation}
			\label{equa:solvesB}
			\! N_b \!\!=\! \frac{1}{2}\mathop{\inf}_{\tau\ge0}\!\left\{K(2M\!+\!\tau^{2})\! + \! (N\!-\!K)\int_{\tau}^\infty\!\!\!(u\!-\!\tau)^2\phi_{2M}(u)\mathrm{d}u\right\}\!\!.\vspace{-2.5ex}
		\end{equation}
	\end{small}
	\noindent The curve indicated in \eqref{equa:solvesB} fits the phase transitions in FAR, as numerically verified in Section ~\ref{section:sim}.
	 
	The tightness of \eqref{equa:solvesB} makes it a powerful tool for guiding waveform design and evaluating recovery performance of \ac{far}. For given system parameters $ M $ and $ N $, as well as $ K $, which means that we have some prior knowledge on the number of targets in a single \ac{crr} bin, \eqref{equa:solvesB} provides the minimum requirement on the number of observations to guarantee unique recovered targets. 
	This is particularly useful when one aims to reduce the number of transmitted pulses $ N_b $ out of $ N $, for the purposes of lowering power consumption \cite{Ma2020joint}, facilitating spectrum sharing between radar and communication \cite{ma2020spatial}, or interference rejection \cite{huang2020multi}. 
For a given tuple of parameters $ (M,N,N_b) $, \eqref{equa:solvesB} implies an equation with respect to $ K $, the maximum number of recoverable targets, evaluating the performance of radars equipped with these parameters. 
This equation with respect to $ K $ can be efficiently solved by iterative methods, e.g., the bisection method \cite{eiger1984bisection}, because $ N_b $ is a monotonic function with respect to $ K $, as stated below.
\begin{mypro}
	\label{pro:mono}
The right hand side of \eqref{equa:solvesB} increases monotonically with $K$.
\end{mypro}
\begin{proof}
	See \ref{proof:increase}.
\end{proof}
	
	
	
	Numerical calculation of \eqref{equa:solvesB} involves complicated integration operation, and when $M$ is a slightly large number, the precise calculation of \eqref{equa:solvesB} is difficult. To avoid the computational burden and allow real-time calculation in practical scenarios, we approximate and simply \eqref{equa:solvesB} under different quantitative relations between $N$ and $K$, as given in the following proposition.
	
	
	
	\begin{mypro}
		\label{pro:appprx nb}
		For large $M$ and different orders of magnitude of $\frac{N}{K}$, $N_b$ in \eqref{equa:solvesB} can be approximated by:
		\begin{itemize}
			\item [i)] when $\frac{N}{K} \gg \sqrt{M}$, 
			\begin{footnotesize}
			\begin{equation}\label{eq:nbcase1 prop}
			\begin{aligned}
			\hspace{-7.5ex}N_b \approx {N}_{b1} = 2MK-\frac{K}{4}+\frac{\sqrt{2}}{2}\cdot K \sqrt{(4M-1)\log\frac{N-K}{K\sqrt{4M-1}}}.
			\end{aligned}
			\end{equation}
			\end{footnotesize}
			\item[ii)] when $\frac{N}{K} \approx \sqrt{M}$,
			\begin{equation}\label{eq:nbcase2 prop}
				N_b \approx {N}_{b2} = KM + \frac{K}{2}x_b(\tau_{\ast}),
			\end{equation}
		where
		\begin{footnotesize}
			\begin{align}
				\hspace{-4ex}x_b(\tau_{\ast}) &=   2M-\frac{3}{4}+\frac{N}{4K}-\frac{(4M-1)K}{N+K}\notag \\
				&\quad +\frac{(N-K)^2}{2K\pi (N+K)}-\frac{(\sqrt{2}+1)(N-K)}{2(N+K)} \sqrt{\frac{4M-1}{\pi}}.    
			\end{align}
		\end{footnotesize}
		\item[iii)] when $\frac{N}{K} \ll \sqrt{M}$, 
		\begin{equation} \label{eq:nbcase3 prop}
				N_b \approx {N}_{b3} = MN-\frac{(4M-1)(N-K)^2}{4N}.
		\end{equation}
		\end{itemize}
	\end{mypro}
	\begin{proof}
		See ~\ref{proof:xb}.
	\end{proof}

We will show in the next section that when $ M \ge 4 $, these approximations are quite accurate. Among these three relations between $ N $ and $ K $, the case of $ N/K \gg \sqrt{M} $ is of particular interest, representing that the observed target scene is relatively sparse. We will compare this curve with the counterpart of standard sparse recovery in the next subsection.
	
	\subsection{Standard Sparse Recovery}
	\label{subsection:far standard}	
	
	To compare \ac{far}'s recovery performance between non-block and block sparse recovery, i.e., \eqref{equa:l1} and \eqref{equa:l21}, we also use \eqref{eq:complexcurve} to indicate the required minimum number of radar echoes, denoted by $N_s$, when applying \eqref{equa:l1}. In this case, the ``block'' size is $m = 1$, and the length of $\bm x$ is $md=d= MN$. The `block' sparsity becomes $s_B = KM$, because each target leads to $M$ nonzero entries in $\bm x$.
	Substituting these variables into \eqref{eq:complexcurve}, we obtain \vspace{-2.5ex}
	
	\begin{equation}
		\label{equa:solvesSR}
		\begin{small}
			\!N_s \!=\! \frac{M}{2} \!\!\mathop{\inf}_{\tau\ge0}\left\{K(2\!+\!\tau^{2})\!\! + \! (N\!\!-\!K)\!\int_{\tau}^\infty\!\!\!\!\!(u\!-\!\tau)^2\phi_{2}(u)\mathrm{d}u\right\}\!\!.\vspace{-0.5ex}
		\end{small}
	\end{equation}
	\noindent Particularly, \eqref{equa:solvesB} and \eqref{equa:solvesSR} are identical when $M=1$.
	
	Similarly to Proposition~\ref{pro:appprx nb}, we have the following proposition that present approximations to \eqref{equa:solvesSR}.
	\begin{mypro}
		\label{pro:xs app}
		When $\frac{N}{K} \gg 1$ and $\frac{N}{K} \approx 1$, \eqref{equa:solvesSR} is approximated by
			\begin{equation}\label{eq:nscase1 prop}
			\begin{aligned}
				N_s \approx N_{s1} =  2MK + \frac{MK\tau_{\star}^2}{2},
				\end{aligned}
			\end{equation}
			and
			\begin{equation}
				N_s \approx N_{s2} = MN-\frac{\pi M(N-K)^2}{4N},
			\end{equation}
respectively, where $ \tau_{\star} $ is the solution of the following equation
\begin{equation}
	\log (\tau_{\star}^2+1) = \log \frac{N-K}{K} -\frac{\tau_{\star}^2}{2}.
\end{equation}
	\end{mypro}
	\begin{proof}
		See \ref{proof:xs}.
	\end{proof}
	\noindent This proposition, like the counterpart for block sparse recovery, intuitively reveals the relationship between $ N_s $ and parameters $ (M,N,K) $, facilitating the comparison between block and standard sparse recovery.

	\subsection{Discussion}
	\label{subsection:far comp}
	
	In the sequel, we discuss the obtained bounds when $ M $ is reasonably large and the observed target scene is relatively sparse, i.e., $ N \gg K $, which occurs in many practical scenarios. Under such conditions, we adopt the approximations \eqref{eq:nbcase1 prop} and \eqref{eq:nscase1 prop}.

	We first note that the obtained conditions that guarantee unique recovery are tighter than the previous counterparts presented in \cite{wang2019theoretical,huang2018analysis}. The previous results, $ K_{b} = O(\frac{N}{M\log(MN)})$ and $ K_{s} =O(\sqrt{\frac{N}{M^2 \log(MN)}})$, (the subscripts denote block and standard sparse recovery, respectively), are based on coherence techniques, which lead to pessimistic bounds. 
	To facilitate the comparison between our results and $ K_{b}$, $ K_{s}$, we set ${N}_{b1} = N$ and $ {N}_{s1} = N $ in \eqref{eq:nbcase1 prop} and \eqref{eq:nscase1 prop}, respectively, because the intact observation models are considered in \cite{wang2019theoretical,huang2018analysis}, where all $ N $ pulses are transmitted, received and processed. 
	Regarding \eqref{eq:nbcase1 prop}, we have $ K_{\rm ib}  \approx \frac{N}{2M + \sqrt{2M} \log \frac{N}{2\sqrt{M}} }$ for intact block sparse recovery (hence the subscript `ib'). Since in practice $ N $ is usually not extremely larger than $ M $, we have ${2M} \ge  \sqrt{2M}\log \frac{N}{2K\sqrt{M}} $. As a consequence, $ K_{\rm ib} $ scales as $ O\left(N/M\right) $, larger than $O(\frac{N}{M \log(MN)})$, indicating the tightness of $ K_{\rm ib} $ over $ K_{b} $. Similarly, from \eqref{eq:nscase1 prop}, we have $ K_{\rm is}  \approx \frac{N}{2M + M \log N}$ for intact standard sparse recovery (hence the subscript `is'), which is simplified as $ K_{\rm is}  =O(\frac{N}{M \log N}) $. In comparison with $ K_s $, we find $ K_{\rm is} $ scales larger than $ K_s $. 	
	We will show by simulations in the next section that the approximations we derive in this section are tight for \ac{far}, while the previous bounds \cite{wang2019theoretical,huang2018analysis} are quite pessimistic.
	
	The tightness enables these approximations to accurately characterize the recovery performance, and facilitates the performance comparison between block and standard sparse recovery for extended targets. 		
	In particular, from \eqref{eq:nbcase1 prop} and \eqref{eq:nscase1 prop}, we have $ N_b = 2MK + O\left( K\sqrt{M \log \frac{N}{K\sqrt{M}}}\right)  $ and $ N_s = 2MK + O\left( KM \log\frac{N}{K}\right)  $, respectively, suggesting $ N_b < N_s $ for reasonably large $ M $. This means that for given $ (M,N,K) $, i.e., under the same system settings and sparse target scene, block sparse recovery requires less observations to guarantee unique recovery of extended targets, implying that block sparse recovery is generally more suitable for recovering extended targets with \ac{far}. \vspace{-2ex}

	\section{Simulation Results}\label{section:sim}
	In this section, simulations are conducted to verify the theoretical curves derived for Gaussian matrices to Section~\ref{section:PT in BSR} and test their application in FAR. 
	In Section \ref{subsection:simPT} and Section \ref{subsection:ph in far}, we measure the success rates of recovering $\bm x$. We assert that $\bm{x}$ is recovered successfully when $\bm{\hat{x}}$, the estimation of $\bm{x}$, satisfies $\|\bm{\hat{x}}-\bm{x}\|_{2} \le 10^{-5}$. 
	In last Section ~\ref{sub:appro}, we examine the proximity of our approximations to the theoretical results.\vspace{-2ex}
	
	\subsection{Phase Transitions under Gaussian Matrices}\label{subsection:simPT}
	This subsection carriers out simulation experiments to inspect the phase transitions on block sparse recovery and the phase transitions in complex-valued Gaussian matrices. 
	
\begin{figure}[h]
		\centering
		\subfigure[Real-valued, block]{
			\includegraphics[width = 0.45\linewidth]{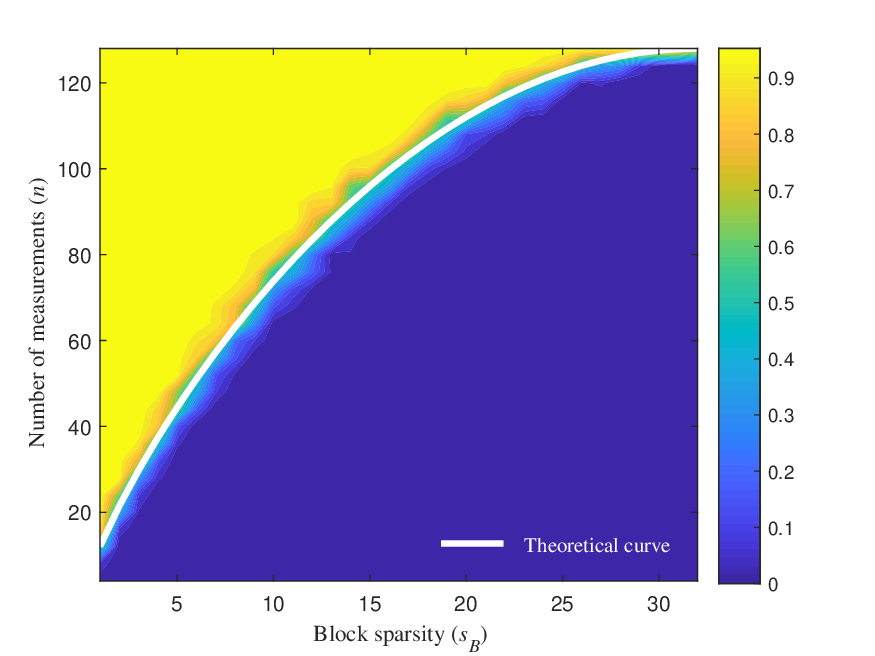}}
		\subfigure[Complex-valued, non-block]{
			\includegraphics[width = 0.45\linewidth]{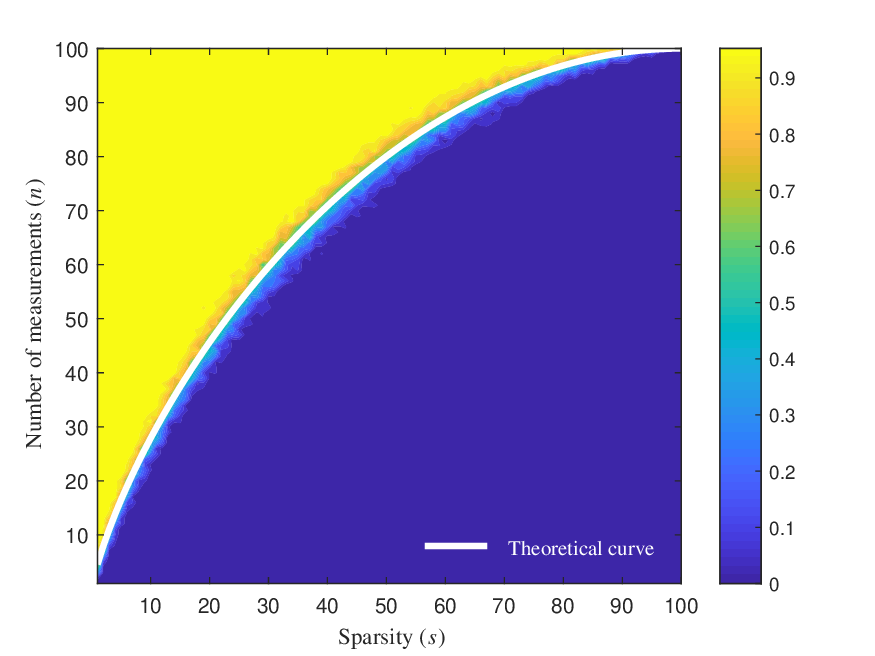}}
		\caption{Phase transitions under real and complex-valued Gaussian matrices using block and non-block sparse recovery, respectively.}
		\label{fig3}
	\end{figure}

	The first experiment considers real-valued block sparse recovery described in \eqref{equa:l21}, where the entries of the observation matrix $\bm{\Psi} \in \mathbb{R}^{n\times md}$ obey i.i.d. $\mathscr{N}(0,1)$. The nonzero entries in $\bm{x}\in\mathbb{R}^{md}$ are 1 or -1 randomly with an identical probability $1/2$. 
	We set $m = 4$ and $d = 32$, 
	and vary $(s_B,n)$ to calculate the probabilities of exact recovery. 
	In the second simulation, we test the phase transition in complex-valued non-block sparse recovery \eqref{equa:l1}, which can be solved by real-valued block sparse recovery as discussed in Subsection~\ref{subsection: PT in CNF}. Here, the entries of $\bm{\Psi} \in \mathbb{C}^{n\times d}$ have their real and imaginary parts obeying i.i.d. $\mathscr{N}(0,1)$. There are $s$ nonzero entries in $\bm{x}\in \mathbb{C}^{d}$, whose phases are i.i.d. $U([0,2\pi])$ and amplitudes equal 1. We set $d = 100$. 
	In both experiments, 50 trials are performed on each pair $(s_B,n)$ or $(s,n)$ to calculate the success rates. The results for these two experiments are shown in Fig.~\ref{fig3} (a) and (b).
	The theoretical curves in (a) and (b) are computed by \eqref{eq:block curve} and \eqref{eq:complexcurve} with corresponding $m$ and $d$, respectively. 
	
	Fig.~\ref{fig3} shows that the theoretical curves conform to empirical phase transitions, which verifies Proposition~\ref{pro:ph in bsr} and \eqref{eq:complexcurve}.\vspace{-2ex}



	\subsection{Phase Transition in FAR Model}\label{subsection:ph in far}
	We next verify existence of phase transitions in FAR. Both standard and block sparse recovery methods are tested.
	
\begin{figure}[h]
		\centering
		\subfigure[Standard sparse recovery]{
			\includegraphics[width = 0.45\linewidth]{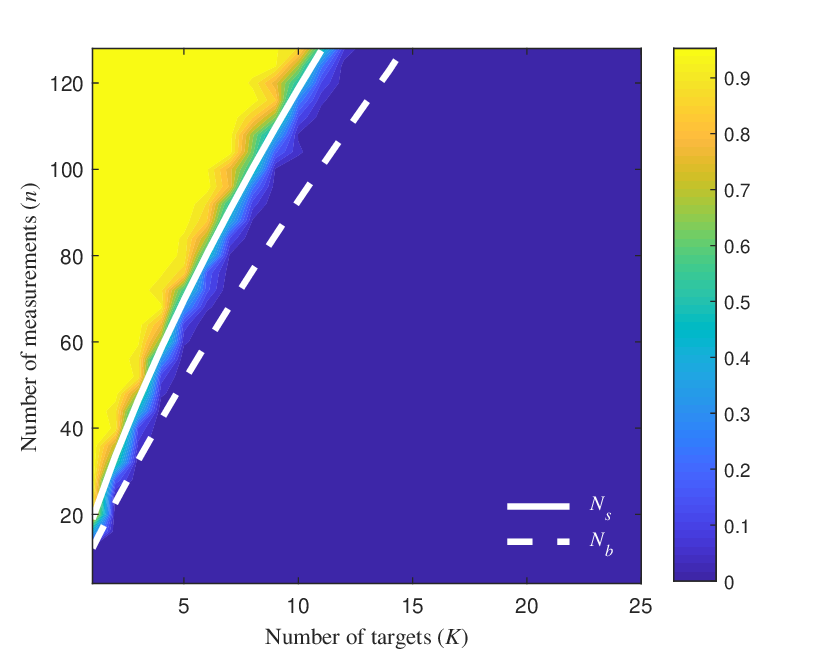}}
		\subfigure[Block sparse recovery]{
			\includegraphics[width = 0.45\linewidth]{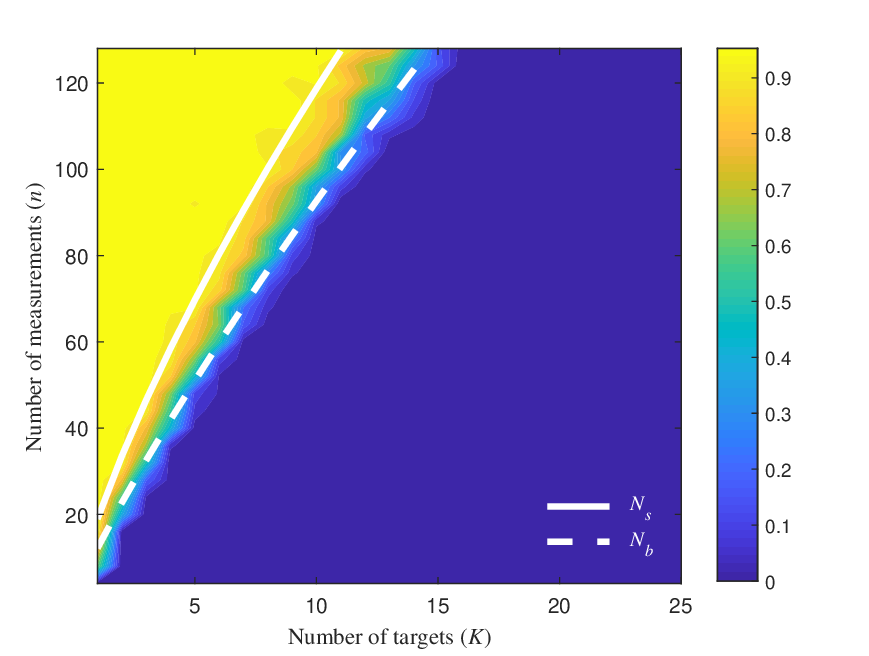}}
		\caption{Phase transitions in \ac{far} using standard and block sparse recovery.\vspace{-4ex}}
		\label{fig6}
	\end{figure}

	To inspect the phase transition in FAR \eqref{equa:FARmodel}, we randomly select $n$ rows from $\bm{\Theta} \in \mathbb{C}^{N\times MN}$ to form a partial measurement matrix $\hat{\bm{\Theta}} \in \mathbb{C}^{n\times MN}$. The phases of nonzero entries in $\bm{x}$ are i.i.d. $U([0,2\pi])$ and the amplitudes equal 1. Given the observations $\bm y =\hat{\bm{\Theta}}\bm{x}$, we use both standard \eqref{equa:l1} and block \eqref{equa:l21} sparse recovery to estimate $\bm{x}$. Recall that in standard sparse recovery, the sparsity is $K M$. We set the parameters $M=4$, $N =128$, $\frac{\Delta f}{f_c}=0.02$, and use 50 trials to calculate the success rates.
	The results of \eqref{equa:l1} and \eqref{equa:l21} are shown in Fig.~\ref{fig6} (a) and (b), respectively. 
	The theoretical curves are calculated with corresponding $M$ and $N$ by \eqref{equa:solvesSR} and \eqref{equa:solvesB}, denoted by `$N_s$' and `$N_b$', representing standard and block sparse recovery, respectively.
	For the sake of comparison between these two sparse recovery methods, we depict both theoretical curves in each figure of phase transition results.
	
	
	From Fig.~\ref{fig6}, we see that both theoretical curves well match their corresponding phase transition curves of FAR. 
	Let $N_s = N_b = N$. The predicted numbers of recoverable
	targets under this setting are $K=11.1$ and $K=14.7$ for standard and block sparse recovery, respectively, whose corresponding success rates are 0.54 ($K=11,\ n=128$ in Fig.~\ref{fig6} (a)) and 0.50 ($K=14,\ n=128$ in Fig.~\ref{fig6} (b)). These rates are close to the threshold $1/2$ that divides the parameter plane into regions of success and failure, indicating that the obtained values of $K$ are tight. 
	However, the counterparts obtained from \cite{huang2018analysis} and \cite{wang2019theoretical} are pessimistically $K=0.4$ and $K=0$, respectively. 
	We also find that the curve of `$N_b$' is generally lower than that of `$N_s$', revealing that block sparse recovery behaves better than standard sparse recovery in the tested cases.\vspace{-2ex} 
	
	
	
	\subsection{Approximation of Phase Transition}\label{sub:appro}
	In this subsection, we will show by simulations that our approximations of the phase transition curves are sufficiently close to the theoretical results, so that we can use them in practical applications to avoid time-consuming calculation.
	
\begin{figure}[h]
		\centering
		\subfigure[$N=128,M=4$]{
			\includegraphics[width = 0.45\linewidth]{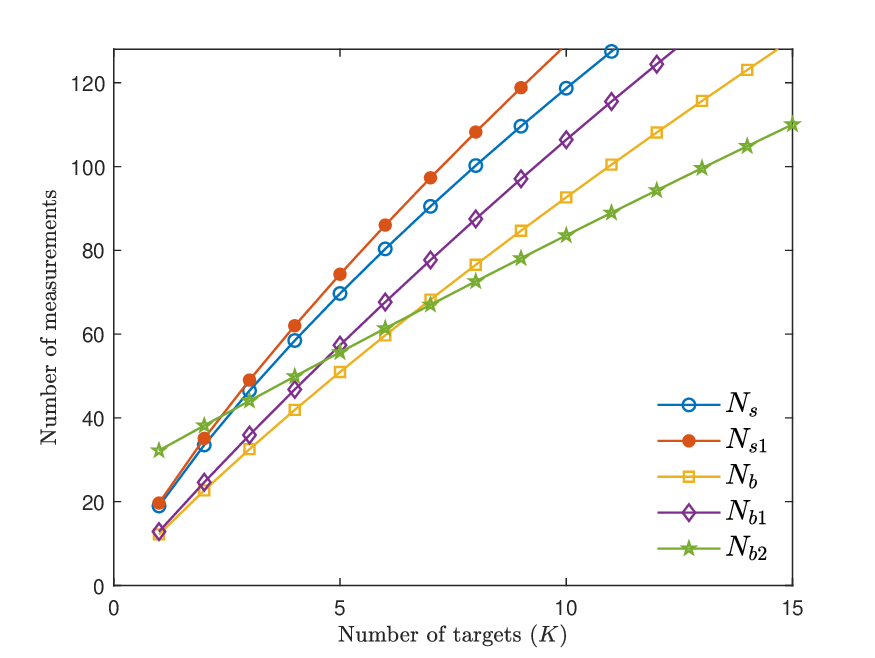}}
		\subfigure[$N=128,M=6$]{
			\includegraphics[width = 0.45\linewidth]{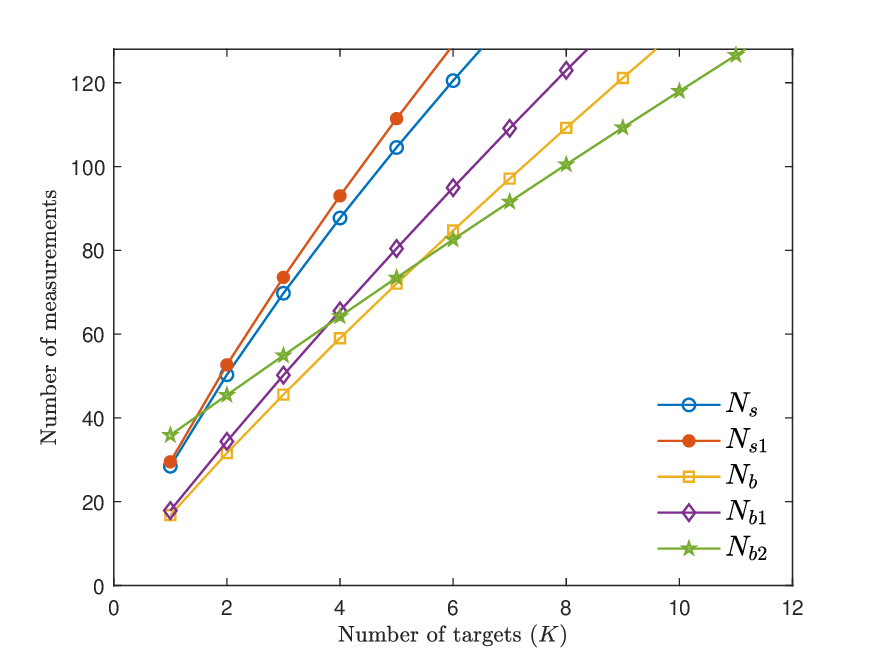}}
		\subfigure[$N=256,M=10$]{
			\includegraphics[width = 0.45\linewidth]{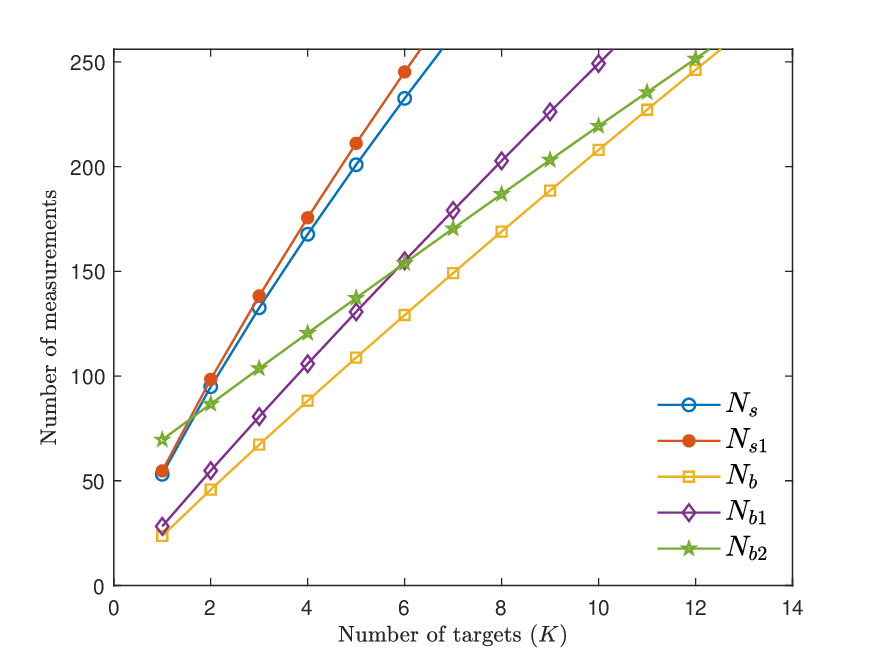}}
		\subfigure[$N=256,M=12$]{
			\includegraphics[width = 0.45\linewidth]{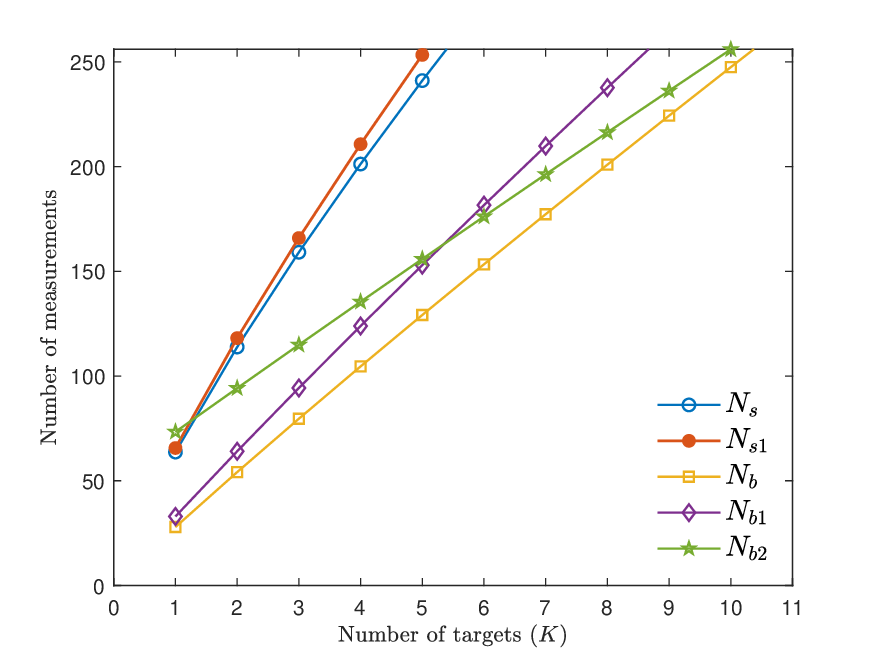}}
		\caption{Comparison between $N_b$, $N_s$, $N_{b1}$, $N_{b2}$ and $N_{s1}$.\vspace{-2ex}}
		\label{fig7}
	\end{figure}

	We compute $N_b$ and $N_s$ versus $ K $ with \eqref{equa:solvesB} and \eqref{equa:solvesSR} to represent the theoretical results under different $N$ and $M$. Then we calculate the values of $N_{b1}$, $N_{b2}$ and $N_{s1}$ in \eqref{eq:nbcase1 prop}, \eqref{eq:nbcase2 prop} and \eqref{eq:nscase1 prop}, respectively, to compare with the theoretical results. We set the parameters $(N,M)$ to be (128,4), (128,6), (256,10) and (256,12), and the corresponding results are shown in Fig.~\ref{fig7} (a-d), respectively. Since $ N_b $ and $ N_s $ cannot exceed $ N $, we restrict their scales between 0 and $N $ in these figures.\vspace{-3ex}
	
	From Fig.~\ref{fig7}, we see that $N_{s1}$ well approximates $N_s$ under all the tested scenarios. As expected in the discussions over Proposition~\ref{pro:appprx nb}, when both $M$ and $K$ are small, $N_{b1}$ is closer to $N_b$ than $N_{b2}$, while $N_{b2}$ behaves better in fitting $N_b$ when either $M$ or $ K $ increases. Fig.~\ref{fig7} also shows that the larger $M$ grows, the more block sparse recovery outperforms the standard counterpart, which is in accordance with our analysis in Section~\ref{subsection:far comp}.\vspace{-2ex}

	\section{Conclusion}\label{section:conclusion}
	In this paper, standard and block sparse recovery for FAR are studied from the perspective of phase transitions. We generalize the phase transitions of standard sparse recovery under real Gaussian matrices to the cases associated with block sparse recovery and complex-valued Gaussian matrices, which numerically conform to phase transitions existing in FAR. 
	We then approximate the obtained phase transition curves with some elementary functions, explicitly revealing the quantitative relationship between the required number of measurements and the numbers of radar pulses, frequencies and targets, as well as facilitating the calculation of these curves. These approximations with analytical expressions are tighter than previous results in \cite{huang2018analysis,wang2019theoretical}, and indicate that block sparse recovery requires less measurements to exactly reconstruct extended targets. 
	Numerical results demonstrate the accuracy of the derived curves and their approximations.\vspace{-2ex}  

	\begin{appendices}
		\renewcommand{\appendixname}{Appendix}
		\renewcommand\thesection{\appendixname~\Alph{section}}

		\section{Proof of Proposition \ref{pro:ph in bsr}}\label{proof:bsr}
		
		In this section, we prove Proposition \ref{pro:ph in bsr}. 
		According to Proposition \ref{thm:upperbound}, we first calculate the subdifferential of the $\ell_{2,1}$ norm defined in \eqref{eq:subdiff}, and then the distance between the subdifferential and a Gaussian vector, as indicated in \eqref{equa:J}. 
		
		Regarding the subdifferential of $\ell_{2,1}$ norm of a $s_B$ block sparse vector $ \bm x \in \mathbb{R}^{md}$, we first reorganize the vector $ \bm x $ for notation convenience. Without loss of generality, we assume that the support set of $ \bm x $ is $\mathcal{B} = \{0, 1,\dots, s_B-1\} $, such that 
		\begin{equation*}
			\bm x^T = \left[\bm x_0^T, \bm x_1^T, \dots, \bm x_{s_B-1}^T, \bm 0^T, \dots, \bm 0^T \right].
		\end{equation*}
		Here, the first $ s_B $ blocks $ \bm x_i \in \mathbb{R}^m$, $ i \in \mathcal{B} $, are nonzero. We use $x_{i,j}$ to represent the $ j $-th element of the $ i $-th block $ \bm x_i $, i.e., the $(im+j)$-th entry of $ \bm x $.  
		Let $ \mathcal{B}^c : = \left\{s_B, s_B+1,\dots, d\right\} $ be the complementary set of $ \mathcal{B} $. 
		
		We link the subdifferential of $ \|\bm{x}\|_{2,1} $ and that of $ \|\bm{x}_i\|_{2} $ by introducing the following lemma. \vspace{-1.5ex}
		\begin{lemma}
			\label{lem:subblock}
			Given two block vectors $\bm x = \left[\bm x_0^T, \bm x_1^T, \dots, \bm x_{d-1}^T \right]^T$, $\bm s = \left[\bm s_0^T, \bm s_1^T, \dots, \bm s_{d-1}^T \right]^T \in \mathbb{R}^{md}$ with $\bm{x}_i $, $\bm{s}_i \in \mathbb{R}^{m}$, the following two statements are equivalent:
			
			1) $\bm{s} \in \partial \|\bm{x}\|_{2,1}$, 
			
			2) $\bm{s}_i \in \partial \|\bm{x}_{i}\|_2, \ i = 0,1,\cdots,d-1$.\vspace{-1.5ex}
		\end{lemma}
		\begin{proof}
			According to \eqref{eq:subdiff}, the definition of subdifferential, we rewrite the above statements into inequalities, respectively:
			
			1) for $\forall \bm{y} \in \mathbb{R}^{md}$, $\|\bm{y}\|_{2,1}\ge \|\bm{x}\|_{2,1} + \langle\bm{y-x},\bm{s} \rangle$,  
			
			2) for $\forall \bm{y}_i \in \mathbb{R}^{d}$, $\|\bm{y}_i\|_{2}\ge \|\bm{x}_i\|_{2} + \langle\bm{y}_i-\bm{x}_i,\bm{s}_i \rangle$, $i = 0,1,\cdots,d-1$.
			
			To prove 2) $ \to $ 1), we set $\bm y = [\bm{y}_0^T ,\bm{y}_1^T ,\cdots,\bm{y}_{d-1}^T ]^T$. By summing both sides of the inequality in 2) with respect to $i$, we have \vspace{-2ex}
			\begin{equation}
				\sum_{i=0}^{d} \|\bm{y}_i\|_{2}  \ge \sum_{i=0}^{d}\|\bm{x}_i\|_{2}  + \sum_{i=0}^{d} \langle\bm{y}_i-\bm{x}_i,\bm{s}_i \rangle,\vspace{-1ex}
			\end{equation}
			where the summation terms are equal to $ \|\bm{y}\|_{2,1} $, $\|\bm{x}\|_{2,1}$ and  $\langle\bm{y-x},\bm{s} \rangle$, respectively, implying 1).
			
			For the other direction 1) $ \to $ 2), we construct $\bm{y} = [\bm{x}_0^T , \cdots,\bm{x}_{i-1}^T , \bm{y}_{i}^T,\bm{x}_{i+1}^T,\cdots,\bm{x}_{d-1}^T ]^T$. Due to the arbitrariness of $\bm{y}$, the inequality in 1) still holds, directly yields the $i$-th inequality in 2) with some simple arrangement, $i = 0,1,\cdots,d-1$. 
			
			Therefore, these two statements are equivalent.			
		\end{proof}		
		
		Lemma \ref{lem:subblock} facilitates the calculation of $ \partial \|\bm{x}\|_{2,1} $: we just need to calculate $ \partial \|\bm{x}_{i}\|_2 $. For a convex function $f(\cdot)$, if it is differentiable at a certain point $\bm{p}$, then the subdifferential of $f(\cdot)$ at $\bm{p}$ contains only one element, the differential of $f(\cdot)$ at $\bm{p}$ \cite[$\S$35]{rockafellar1970convex}.
		Therefore, we discuss $ \partial \|\bm{x}_i\|_{2} $ when $ \bm x_i $ is a nonzero or zero block, respectively.  
		In the former case, the function $ \|\bm{x}_i\|_{2}  $ is differentiable with respective to $ \bm x_i $. The partial differential of $ \|\bm{x}_i\|_{2} $ with respective to an entry $ x_{i,j} $ is given by
		\begin{equation}
			\label{eq:diff at nonzero}
			\frac{x_{i,j}}{\sqrt{\sum_{k=0}^{m-1}x_{i,k}^2}}, i \in \mathcal{B} ,j \in \mathcal{M}.
		\end{equation}
		In the latter case, the partial differential of $ \|\bm{x}_p\|_{2} $,  $  p \in \mathcal{B}^{c}$, does not exist, and we will calculate its subdifferential with the following lemma. 
		
		\begin{lemma}
			\label{lem:subdiff l2}
			Given the function $f(\bm{x}) = \|\bm{x}\|_2$, $\bm{x}\in \mathbb{R}^{d}$, the subdifferential of $f(\cdot)$ at $\bm{x} = \bm{0}$ is $\{\bm{s}\in \mathbb{R}^{d}: \|\bm{s}\|_2 \le 1 \}$.
		\end{lemma}
		\begin{proof}
			Let $\mathcal{S} = \{\bm{s}\in \mathbb{R}^{d}: \|\bm{s}\|_2 \le 1 \}$. To prove $\mathcal{S} = \partial f(\bm{0})$, we need to show that $\mathcal{S} \subset \partial f(\bm{0})$ and $\partial f(\bm{0}) \subset \mathcal{S}$.
			
			We first consider $\mathcal{S} \subset \partial f(\bm{0})$. Recall the definition of the subdifferential, $\partial {f}(\bm{x})\coloneqq \{\bm s\in\mathbb{R}^d: \\ f(\bm{y})\ge f(\bm{x})+\langle\bm{s},\bm{y}-\bm{x}\rangle, \ \forall \  \bm{y}\in\mathbb{R}^d\}$. For $\forall \bm{s} \in \mathcal{S}$, using the Cauchy-Buniakowsky-Schwarz inequality, we have the following inequality\vspace{-1ex}
			\begin{equation}
				\|\bm{0}\|_2 + \langle\bm{s},\bm{y}-\bm{0}\rangle \le {\|\bm{s}\|_2 \|\bm{y}\|_2} \le  \|\bm{y}\|_2,
			\end{equation}
			being true for $\forall \bm{y} \in \mathbb{R}^{d}$, which proves $\mathcal{S} \subset \partial f(\bm{0})$.
			
			We then state $\partial f(\bm{0}) \subset \mathcal{S}$ by proving its contrapositive: for $\forall \bm{s} \in \mathbb{R}^{d}$, if $\bm{s} \not\in \mathcal{S}$, then $\bm{s} \not \in \partial f(\bm{0})$. For $\bm{s} \in \mathbb{R}^{d}$, $\|\bm{s}\|_2 > 1$, we choose $\bm{y} = \frac{\bm{s}}{\|\bm{s}\|_2}$, and obtain the following inequality
			\begin{equation}
				f(\bm{y}) = \|\bm{y}\|_2 = 1 < \|\bm{s}\|_2 = \langle\bm{y},\bm{s}\rangle = \langle\bm{y}-\bm{0},\bm{s}\rangle + f(\bm{0}),
			\end{equation}
			implying $\bm{s} \not\in \partial f(\bm{0})$.
			
			With the two parts above, we prove $\partial f(\bm{0}) = \mathcal{S}$.
		\end{proof}
	\noindent
		Lemma \ref{lem:subdiff l2} completes the subdifferential of $ \|\bm x_p\|_2 $ for $ p \in \mathcal{B}^c $. Recall that when $ i \in \mathcal{B} $ the subdifferential is given by  \eqref{eq:diff at nonzero}.
		
		We are now ready to derive $  \partial \|\bm{x}\|_{2,1} $. Let $ \bm v \in \mathbb{R}^{md} $  denote an element in the subdifferential of $ \|\bm{x}\|_{2,1} $, i.e., $ \bm{v}\in \partial \|\bm{x}\|_{2,1} $, with its $ i $-th block denoted by $ \bm v_i \in \mathbb{R}^m $ and $(im+j)$-th entry by $v_{i,j}$. The subdifferential $  \partial \|\bm{x}\|_{2,1} $ forms a cone, given by \vspace{-1ex}
		\begin{small}
		\begin{align}
			\partial \|\bm{x}\|_{2,1}= \Bigg\{ 
			\bm v \in \mathbb{R}^{md}: &
			v_{i,j} = \frac{x_{i,j}}{\sqrt{\sum_{k=0}^{m-1}x_{i,k}^2}},
			\sum_{q=0}^{m-1}v_{p,q}^2\le 1,   \notag \\
			& i \in \mathcal{B},  j \in \mathcal{M}, p \in \mathcal{B}^c
			\Bigg\}.\vspace{-5ex}
			\label{eq:subdiff of l21}
		\end{align}
		\end{small}

		We then calculate the distance between a standard normal vector $\bm{g} \in \mathbb{R}^{md}$, $\bm g \sim \mathscr{N}(0,\bm I_{md})  $, and the set $  \partial \|\bm{x}\|_{2,1} $. 
		Let $\bm g_{i}\in \mathbb{R}^m$ and $g_{i,j}$ represent the $ i $-th block and $(im+j)$th entry in $\bm{g}$, respectively. The distance is then calculated block-wise, given by\vspace{-2ex}
		\begin{equation}
			{\rm dist}^{2}\left(\bm{g},\tau \cdot \{ {\bm v} \}\right) = \sum_{i=0}^{d-1} {\rm dist}^{2}\left(\bm{g}_i,\tau \cdot \{ {\bm v}_i \}\right).\vspace{-1ex}
		\end{equation}
	For $ i \in \mathcal{B} $, $ {\rm dist}^{2}\left(\bm{g}_i,\tau \cdot  {\bm v}_i \right) = \sum_{j=0}^{m-1}(g_{i,j}-\tau v_{i,j})^2 $, because the subdifferential reduces to a single point. When $ p \in \mathcal{B}^c $, we have $ \{\bm v_p\} = \{\bm s \in \mathbb{R}^m: \|\bm s\|_2 \le 1\} $. Therefore, the distance is given by\vspace{-2ex}
		\begin{equation} 
				{\rm dist}^{2}\left(\bm{g}_pu \cdot \{ {\bm v}_p \}\right) = \inf_{\|\bm v_p\|_2 \le 1} \sum_{q=0}^{m-1}(g_{p,q}-\tau v_{p,q})^2, 
		\end{equation}
		which equals zero when $ \|\bm g_p\|_2 \le \tau $ and $\left( \|\bm g_p\|_2 - \tau\right)^2 $ otherwise, hence is expressed by $G_p(\tau):= \max((\|\bm g_p\|_2 - \tau)^2,0) $. 
	From the above discussion, we have 
	\begin{footnotesize}
\begin{align} \label{eq:dist2}
		{\rm dist}^{2}\left(\bm{g},\tau \cdot \{ {\bm v} \}\right) = \sum_{i=0}^{s_B-1}\sum_{j=0}^{m-1}(g_{i,j}-\tau v_{i,j})^2 
		+\sum_{p=s_B}^{d-1}G_p(\tau). 
\end{align}	
\end{footnotesize}
		
		We next calculate the expectation.
		We have $\mathbb{E}\left[g_{i,j}^2\right] = 1$, $\mathbb{E}\left[g_{i,j}v_{i,j}\right]=0$ and $\sum_{j=0}^{m-1} v_{i,j}^2 = 1,\ i\in\mathcal{B}$, $ j \in \mathcal{M} $. The expectation of the first term in \eqref{eq:dist2} is rewritten as follows
		\begin{small}
		\begin{align}
			&\mathbb{E} \left[\sum_{i=0}^{s_B-1}\sum_{j=0}^{m-1}(g_{i,j}-\tau v_{i,j})^2\right] = \mathbb{E} \left[ \sum_{i=0}^{s_B-1}\sum_{j=0}^{m-1}g_{i,j}^2\right] \notag\\
			&\quad \quad - 2\tau\mathbb{E} \left[ \sum_{i=0}^{s_B-1}\sum_{j=0}^{m-1} g_{i,j} v_{i,j}\right] + \tau^{2}\sum_{i=0}^{s_B-1}\sum_{j=0}^{m-1} v_{i,j}^2 \notag\\
			&\quad = s_B\left(m+\tau^2\right).
		\end{align}
		\end{small}
		Let $u_p = \|\bm g_p\|_2 $, which obeys the $\chi$-distribution with $m$ degrees of freedom. The expectation of the second term can be calculated as
		\begin{align}
			\mathbb{E}&\left[ \sum_{i=s_B}^{d-1}G_p(\tau)\right]=(d-s_B)  \mathbb{E}\left[{\rm max}((u_p-\tau)^2,0)\right]\notag \\
			&=(d-s_B)\int_{\tau}^\infty(u-\tau)^2\phi_m(u)\mathrm{d}u\notag \\
			&\ \ \ \ + (d-s_B)\int_{0}^{\tau}0 \cdot \phi_m(u)\mathrm{d}u \notag \\
			& = (d-s_B)\int_{\tau}^\infty(u-\tau)^2\phi_m(u)\mathrm{d}u,
		\end{align}
		where $\phi_m(u)$ is the probability density function of the $\chi$-distribution with $m$ degrees of freedom.
		With the above results, the expectation is given by
		\begin{align}
			\mathbb{E}&\left[ {\rm dist}^{2}\left(\bm{g},\tau \cdot \{\tilde{\bm v}\} \right)\right] \notag \\
			&= s_B\left(m+\tau^{2}\right)+(d-s_B)\int_{\tau}^\infty(u-\tau)^2\phi_m(u)\mathrm{d}u.
			\label{eq:e dist}
		\end{align}

		According to Proposition \ref{thm:upperbound}, 
		we obtain the upper bound on $ \delta(D(\left\| \cdot \right\|_{2,1}, \bm x)) $ as 
		\begin{small}
		\begin{align}
			&\varphi_m(s_B,d) =\inf_{\tau\ge 0}\mathbb{E}[ {\rm dist}^{2}(\bm{g},\tau \cdot \partial\|\bm{x}\|_{2,1})] \notag \\
			&= \mathop{\inf}_{\tau\ge0}\{s_B(m+\tau^{2})+(d-s_B)\int_{\tau}^\infty(u-\tau)^2\phi_m(u)\mathrm{d}u\},
		\end{align}
		\end{small}
		which completes the proof.\vspace{-3ex}
		
		\section{Proof of \Revise{Proposition \ref{pro:mono}}}\label{proof:increase}
		
		Here, we prove the monotonicity of $N_b$ in \eqref{equa:solvesB} with respective to $ K $. For convenience, we define 
		$H(K,\tau)$ as
		\begin{align}
			H(K,\tau) := &K(2M+\tau^{2})\notag \\
			&+  (N-K)\int_{\tau}^\infty(u-\tau)^2\phi_{2M}(u)\mathrm{d}u,
		\end{align}
		where $\tau \ge 0$, such that 
		\begin{equation}
			N_b = \frac{1}{2} \inf_{\tau\ge 0} H(K,\tau). 
		\end{equation}

		To reveal the monotonicity, we regard the integer $ K $ as a real number, and calculate the partial derivative of $H(K,\tau)$ with respect to $K$, given by
		\begin{equation}
			\frac{\partial H(K,\tau)}{\partial K} = 2M+\tau^{2} - \int_{\tau}^\infty(u-\tau)^2\phi_{2M}(u)\mathrm{d}u,
		\end{equation}
		which is non-negative as shown below.
		
		Since $ (u-\tau)^2 \le u^2 $ for $ u \ge \tau \ge 0 $ and $ \phi_{2M}(u) \ge 0$ for $ u \ge 0 $, we have 
		\begin{small}
		\begin{align}
			\int_{\tau}^\infty(u-\tau)^2\phi_{2M}(u)\mathrm{d}u &\le \int_{\tau}^\infty u^2\phi_{2M}(u)\mathrm{d}u \notag \\
			& \le \int_{0}^\infty u^2\phi_{2M}(u)\mathrm{d}u,
		\end{align}
		\end{small}
		which leads to 
		\begin{small}
		\begin{align}
			\frac{\partial H(K,\tau)}{\partial K} &\ge 2M+\tau^{2} - \int_{\tau}^\infty u^2\phi_{2M}(u)\mathrm{d}u \notag \\
			&\ge 2M+\tau^{2} - \int_{0}^\infty u^2\phi_{2M}(u)\mathrm{d}u.
			\label{eq:H partial K}
		\end{align}
		\end{small}
		Recall that $ \phi_{2M}(u) $ denotes the probability density function of the $ \chi $-distribution with $ 2M $ degrees of freedom. The integral term in \eqref{eq:H partial K} represents the second moment of $ \chi $-distribution, which equals the first moment of $ \chi^2 $-distribution, i.e., the degrees of freedom $ 2M $ \cite{chung2001course}. Therefore, \eqref{eq:H partial K} results in 
		\begin{equation}
			\begin{aligned}
				\frac{\partial H(K,\tau)}{\partial K} \ge 2M+\tau^{2}-2M= {\tau^{2}} \ge 0.
			\end{aligned}
		\end{equation}
		As a consequence, $H(K,\tau)$ is monotonically non-decreasing with the increase of $ K $, which proves $ H(K_1,\tau) \ge H(K_2,\tau) $, when $K_1 >= K_2$. Take the infimum of $\tau$ on both sides of the inequality, we have $N_b(K_1) >= N_b(K_2)$, completing the proof. \vspace{-2ex}

		\section{Proof of Proposition \ref{pro:appprx nb}} \label{proof:xb}
		To simplify $N_b$ in \eqref{equa:solvesB}, we first introduce 
		\begin{equation}
			\label{eq:xb}
			x_b(\tau):=\tau^2 + \frac{N-K}{K}\int_{\tau}^\infty\!\!\!(u\!-\!\tau)^2\phi_{2M}(u)\mathrm{d}u,
		\end{equation}
		such that  
		\begin{equation}
			N_b = KM + \frac{K}{2} \inf_{\tau\ge 0} x_b(\tau). 
			\label{eq:Nb tau}
		\end{equation}
		In the following, we approximate $ x_b(\tau) $ with a conciser form in \ref{subsec:app xb}. We then seek the value of $ \tau \ge 0$ that leads to the infimum of $ x_b(\tau) $ in  \ref{subsec:inf xb}, followed by the calculation of $ \inf_{\tau\ge 0} x_b(\tau) $ in  \ref{subsec:app inf xb}. 
		Among the derivatives, some formulas corresponding to the integral over a normal distribution will be used \cite{chung2001course}, given by\vspace{-1ex}
		\begin{small}
		\begin{equation}
			\int_{y}^{\infty} u^3 e^{-\frac{u^2}{2}}\mathrm{d}u =  y^2 e^{-\frac{y^2}{2}} + e^{-\frac{y^2}{2}},\vspace{-2ex}
			\label{eq:normal int4}
		\end{equation}
		\begin{equation}
			\int_{y}^{\infty} u^2 e^{-\frac{u^2}{2}}\mathrm{d}u = \sqrt{\frac{\pi}{2}} {\rm erfc}\left(\frac{y}{\sqrt{2}}\right) + y e^{-\frac{y^2}{2}},\vspace{-2ex}
			\label{eq:normal int1}
		\end{equation}
		\begin{equation}
			\int_{y}^{\infty} u e^{-\frac{u^2}{2}}\mathrm{d}u = e^{-\frac{y^2}{2}},\vspace{-2ex}
			\label{eq:normal int2}
		\end{equation}
		\begin{equation}
			\int_{y}^{\infty}  e^{-\frac{u^2}{2}}\mathrm{d}u = \sqrt{\frac{\pi}{2}} {\rm erfc}\left(\frac{y}{\sqrt{2}}\right),\vspace{-2ex}
			\label{eq:normal int3}
		\end{equation}
		\end{small}
		where $ {\rm erfc}(x) = \frac{2}{\sqrt{\pi}}\int^{\infty}_{x} e^{-\eta^{2} d\eta}$ is the complementary error function.\vspace{-2ex} 
		
		\subsection{Approximation of $ x_b(\tau) $}
		\label{subsec:app xb}
		
		The approximation of $ x_b(\tau) $ is based on the central limit theorem \cite{chung2001course}, indicating that the $ \chi $-distribution probability density function $\phi_{2M}(u)$ can be well approximated by a probability density function of a normal distribution when $M$ is reasonably large. Particularly, \vspace{-1ex} 
		\begin{equation}
			\phi_{2M}(u) \approx {\frac {1}{\sigma_M {\sqrt {2\pi }}}}e^{-{\frac {1}{2}}\left({\frac {u-\mu_M }{\sigma_M }}\right)^{2}},\vspace{-1ex} 
			\label{eq:app phi with normal}
		\end{equation}
		where the mean and variance are denoted by $ \mu_M $ and $\sigma_M^2$, respectively, given by\vspace{-2ex} 
		\begin{small}
		\begin{equation}
			\mu_M  = \frac{\sqrt{2}\Gamma(M+1/2)}{\Gamma(M)},\vspace{-2ex} 
			\label{eq:muM}
		\end{equation}
		
		\begin{equation}
			\sigma_M^2  = 2M - \mu_M^2.\vspace{-1.5ex} 
			\label{eq:sigmaM}
		\end{equation}
		\end{small}
		We note that for sufficiently large  $ M $, the mean and variance in \eqref{eq:muM} and \eqref{eq:sigmaM} lead to
		\begin{small}
		\begin{equation}
			\mu_M^2 \approx 2M-1/2,
			\label{eq:muM2 app}
		\end{equation}
		\begin{equation}
			\sigma_M^2 \approx 1/2,
			\label{eq:sigmaM app}
		\end{equation}
		\end{small}
		respectively \cite{Buric:2011:BPA}. In this case, $ \mu_{M}/ \sigma_M = O(\sqrt{M}) $. 
		
		By substituting \eqref{eq:app phi with normal} into \eqref{eq:xb}, we approximate the integral term in the right hand side of \eqref{eq:xb} by
		\begin{footnotesize}
		\begin{equation}
			\int_{\tau}^{\infty} (u-\tau)^2 \phi_{2M}(u) \mathrm{d}u 
			\approx {\frac {1}{\sigma_M {\sqrt {2\pi }}}}\int_{\tau}^{\infty} (u-\tau)^2 e^{-\frac{(u-\mu_M)^2}{2\sigma_M^2}}\mathrm{d}u.
		\end{equation}
		\end{footnotesize}
		Now, 
		\begin{small}
		\begin{align}
			& \int_{\tau}^{\infty} (u-\tau)^2 e^{-\frac{(u-\mu_M)^2}{2\sigma_M^2}}\mathrm{d}u \notag \\
			&= \int_{\tau}^{\infty} (u-\mu_M+\mu_M-\tau)^2 e^{-\frac{(u-\mu_M)^2}{2\sigma_M^2}}\mathrm{d}u,
		\end{align}
		\end{small}
		which can be expanded into a summation of three terms
		\begin{small}
		\begin{align}
			\label{eq:threeterms}
			&\int_{\tau}^{\infty} (u-\mu_M)^2 e^{-\frac{(u-\mu_M)^2}{2\sigma_M^2}}\mathrm{d}u \notag\\
			&\quad + 2(\mu_M-\tau)\int_{\tau}^{\infty} (u-\mu_M) e^{-\frac{(u-\mu_M)^2}{2\sigma_M^2}}\mathrm{d}u \notag\\
			&\quad + \int_{\tau}^{\infty} (\mu_M-\tau)^2 e^{-\frac{(u-\mu_M)^2}{2\sigma_M^2}}\mathrm{d}u.
		\end{align}
		\end{small}
		We next calculate these terms by using formulas \eqref{eq:normal int1}-\eqref{eq:normal int3} individually. The first term in \eqref{eq:threeterms} can be rewritten as
		\begin{small}
		\begin{align}
			\label{eq:squareterms}
			&\int_{\tau}^{\infty} (u-\mu_M)^2 e^{-\frac{(u-\mu_M)^2}{2\sigma_M^2}}\mathrm{d}u \notag\\
			&= \sigma_M^2\int_{\tau}^{\infty} \frac{(u-\mu_M)^2}{\sigma_M^2} e^{-\frac{(u-\mu_M)^2}{2\sigma_M^2}}\mathrm{d}u \notag\\
			&\stackrel{(a)}{=}\sigma_M^3 \cdot
			\left(\sqrt{\frac{\pi}{2}}{\rm erfc}\left(\frac{\tau-\mu_M}{\sqrt{2}\sigma_M}\right) + \frac{\tau-\mu_M}{\sigma_M} e^{-\frac{(\tau-\mu_M)^2}{2\sigma_M^2} }\right),
		\end{align}	 
		\end{small}
		where (a) is a consequence of \eqref{eq:normal int1}. 
		The integral in the second term of \eqref{eq:threeterms} can be simplified by  \eqref{eq:normal int2}, implying
		\begin{small}
		\begin{align}
			\label{eq:linearterms}
			\int_{\tau}^{\infty} (u-\mu_M) e^{-\frac{(u-\mu_M)^2}{2\sigma_M^2}}\mathrm{d}u
			&= \sigma_M\int_{\tau}^{\infty} \frac{u-\mu_M}{\sigma_M} e^{-\frac{(u-\mu_M)^2}{2\sigma_M^2}}\mathrm{d}u \notag\\
			&=\sigma_M \cdot
			\sigma_M e^{-\frac{(\tau-\mu_M)^2}{2\sigma_M^2} }.
		\end{align}	 	 
		\end{small}
		With \eqref{eq:normal int3}, we rewrite the integral of the third term in \eqref{eq:threeterms} as
		\begin{align}
			\label{eq:thirdterms}
			\int_{\tau}^{\infty}  e^{-\frac{(u-\mu_M)^2}{2\sigma_M^2}}\mathrm{d}u 
			=  \sigma_M\sqrt{\frac{\pi}{2}}{\rm erfc}\left(\frac{\tau-\mu_M}{\sqrt{2}\sigma_M}\right) 
			.
		\end{align}
		
		Substituting \eqref{eq:squareterms}-\eqref{eq:thirdterms} into \eqref{eq:threeterms} yields
		\begin{small}
		\begin{align}
			&\int_{\tau}^{\infty} (u-\tau)^2 e^{-\frac{(u-\mu_M)^2}{2\sigma_M^2}}\mathrm{d}u \notag\\
			&= \left(	\sigma_M^3 +(\mu_M-\tau)^2 \sigma_M \right) \sqrt{\frac{\pi}{2}}{\rm erfc}\left(\frac{\tau-\mu_M}{\sqrt{2}\sigma_M} \right)\notag \\  
			&\quad +\left(\sigma_M^2 (\tau-\mu_M) + 2 (\mu_M-\tau)\sigma_M^2 \right)  e^{-\frac{(\tau-\mu_M)^2}{2\sigma_M^2} } \notag \\
			&= \left(	\sigma_M^3 +(\mu_M-\tau)^2 \sigma_M \right) \sqrt{\frac{\pi}{2}}{\rm erfc}\left(\frac{\tau-\mu_M}{\sqrt{2}\sigma_M} \right) \notag\\ 
			&\quad + (\mu_M-\tau)\sigma_M^2   e^{-\frac{(\tau-\mu_M)^2}{2\sigma_M^2} }
			. 
		\end{align}	
		\end{small}
		Plugging this integral into \eqref{eq:xb}, yields
		\begin{small}
		\begin{align}
			x_b(\tau) &= \tau^2 + \frac{N-K}{K\sigma_M {\sqrt {2\pi }}} \int_{\tau}^{\infty} (u-\tau)^2 e^{-\frac{(u-\mu_M)^2}{2\sigma_M^2}}\mathrm{d}u
			\notag \\
			&= \tau^2 + \frac{N-K}{2K} \left(	\sigma_M^2 +(\mu_M-\tau)^2  \right) {\rm erfc}\left(\frac{\tau-\mu_M}{\sqrt{2}\sigma_M} \right) \notag \\
			&\quad + \frac{N-K}{K {\sqrt {2\pi }}} (\mu_M-\tau)\sigma_M  e^{-\frac{(\tau-\mu_M)^2}{2\sigma_M^2} }\vspace{-2ex}
			. 
			\label{eq:xbdetail}
		\end{align}	
		\end{small}
		
		\subsection{Minimizer of $ x_b(\tau) $}
		\label{subsec:inf xb}
		We denote by $ \tau_{\ast} $ the minimizer of $ x_b(\tau) $. Taking partial derivatives over both sides of \eqref{eq:xb} and letting $ \frac{\partial x_b(\tau) }{\partial \tau} = 0 $, we have
		\begin{equation}
			\label{eq:kkt}
			\tau_{\ast} = \frac{N-K}{K}\int_{\tau_{\ast}}^{\infty} (u-\tau_{\ast})\phi_{2M}(u)\mathrm{d}u.
		\end{equation}
		In this subsection, we simplify the integral function \eqref{eq:kkt}, and then substitute the result into \eqref{eq:xbdetail} to facilitate the calculation of $ x_b(\tau_{\ast}) $.
		
		Using \eqref{eq:app phi with normal}, we approximate the integral in \eqref{eq:kkt} by
		\begin{small}
		\begin{align}
			&\int_{\tau_{\ast}}^{\infty} (u-\tau_{\ast}) e^{-\frac{(u-\mu_M)^2}{2\sigma_M^2}}\mathrm{d}u \notag
			\\
			&= \int_{\tau_{\ast}}^{\infty} (u-\mu_M+\mu_M-\tau_{\ast}) e^{-\frac{(u-\mu_M)^2}{2\sigma_M^2}}\mathrm{d}u \notag
			\\
			&=\sigma_M^2 e^{-\frac{(\tau_{\ast}-\mu_M)^2}{2\sigma_M^2} }
			+(\mu_M-\tau_{\ast}) \sigma_M\sqrt{\frac{\pi}{2}}{\rm erfc}\left(\frac{\tau_{\ast}-\mu_M}{\sqrt{2}\sigma_M}\right),
		\end{align}
		\end{small}
	    from \eqref{eq:linearterms} and \eqref{eq:thirdterms}.
		Hence, we rewrite \eqref{eq:kkt} as
		\begin{small}
		\begin{align}
			\tau_{\ast} &=  \frac{N-K}{K\sigma_M {\sqrt {2\pi }}} \int_{\tau_{\ast}}^{\infty} (u-\tau_{\ast}) e^{-\frac{(u-\mu_M)^2}{2\sigma_M^2}}\mathscr{d}u
			\notag \\
			&= \frac{N-K}{2K} (\mu_M-\tau_{\ast}) {\rm erfc}\left(\frac{\tau_{\ast}-\mu_M}{\sqrt{2}\sigma_M} \right) \notag\\
			&\quad + \frac{N-K}{K {\sqrt {2\pi }}} \sigma_M  e^{-\frac{(\tau_{\ast}-\mu_M)^2}{2\sigma_M^2} }
			. 
			\label{eq:taukkt}
		\end{align}	
		\end{small}

		Denote the solution to \eqref{eq:taukkt} by $ \tau_{\ast} = \mu_M - \alpha \sigma_M $, where $ \alpha  $ is unknown indicating the normalized difference between $ \tau_{\ast} $ and $ \mu_M $. Substituting $ \tau_{\ast} = \mu_M - \alpha \sigma_M $ into \eqref{eq:taukkt}, we obtain
		\begin{footnotesize}
		\begin{align}
			\mu_M - \alpha \sigma_M = \frac{N-K}{2K} \alpha \sigma_M {\rm erfc}\left(-\frac{\alpha}{\sqrt{2}} \right) 
			+ \frac{N-K}{K {\sqrt {2\pi }}} \sigma_M  e^{-\frac{\alpha^2}{2} }
			. 
			\label{eq:taukktwrtalpha1}
		\end{align}
		\end{footnotesize}
		After some arrangement, \eqref{eq:taukktwrtalpha1} leads to 
		\begin{small}
		\begin{align}
			&\alpha -\left( \frac{\alpha}{2}  {\rm erfc}\left(-\frac{\alpha}{\sqrt{2}} \right) 
			+ \frac{1}{{\sqrt {2\pi }}}   e^{-\frac{\alpha^2}{2} }\right) \notag\\ &= \frac{\mu_M}{\sigma_M} - \frac{N}{K}\left( \frac{\alpha}{2}  {\rm erfc}\left(-\frac{\alpha}{\sqrt{2}} \right) 
			+ \frac{1}{{\sqrt {2\pi }}}   e^{-\frac{\alpha^2}{2} }\right)
			,
			\label{eq:taukktwrtalpha2}
		\end{align}
		\end{small}
		which can be rewritten as
		\begin{small}
		\begin{align}
			\frac{\alpha}{2}  {\rm erfc}\left(-\frac{\alpha}{\sqrt{2}} \right) 
			+ \frac{1}{{\sqrt {2\pi }}}   e^{-\frac{\alpha^2}{2} } = \left(\frac{\mu_M}{\sigma_M}-\alpha\right) \cdot \frac{K}{N-K}
			. 
			\label{eq:alphacase2}
		\end{align}
		\end{small}
		Note that the function on the left hand side is monotonically increasing with respect to $ \alpha $ while the one on the right hand side is monotonically decreasing. Therefore, there is only one solution to \eqref{eq:alphacase2}. 
		We will substitute the values of $ \alpha $, i.e., $ \alpha \gg 1 $, $ \alpha \approx 0 $ and $ \alpha \ll -1 $, into \eqref{eq:taukktwrtalpha2}, respectively, in order to check whether \eqref{eq:taukktwrtalpha2} holds and reveal the dependency between $ \alpha $ and $ (K, N, M) $. 
		We resort to series expansions of $ {\rm erfc}(\cdot) $ at $ \pm \infty $, respectively, given by
		\begin{small}
		\begin{equation}
			\left. \frac{\alpha}{2} {\rm erfc}\left(\frac{-\alpha}{\sqrt{2}}\right) \right|_{ \alpha = + \infty} 
			\approx \alpha -  \frac{1}{{\sqrt {2\pi }}}   e^{-\frac{\alpha^2}{2} } ,
			\label{eq:xerfc +}
		\end{equation}
		\begin{equation}
			\left. \frac{\alpha}{2} {\rm erfc}\left(\frac{-\alpha}{\sqrt{2}}\right) \right|_{ \alpha = - \infty} 
			\approx  -  \frac{1}{{\sqrt {2\pi }}}   e^{-\frac{\alpha^2}{2} } (1 - 1/\alpha^2 ).
			\label{eq:xerfc -}
		\end{equation}
		\end{small}

		i) We first consider the case $ \alpha \ll -1 $. Substitute \eqref{eq:xerfc -} into \eqref{eq:taukktwrtalpha2}, leads to \vspace{-2ex}
		\begin{equation}
			\alpha \approx \frac{\mu_M}{\sigma_M} - \frac{N - K}{K} \frac{e^{-\frac{\alpha^2}{2} }}{{\sqrt {2\pi }} \alpha^2} ,\vspace{-1ex}
			\label{eq:a case2}
		\end{equation}
		which indicates that satisfying $ \alpha \ll -1 $ requires $ \frac{N}{K} \gg \frac{\mu_M}{\sigma_M}$. 
		After arrangement of \eqref{eq:a case2}, we take logarithm on both sides of approximation, resulting in
		\begin{equation}
			\log \frac{N-K}{K} - \frac{\alpha^2}{2} - \log \sqrt{2\pi}\alpha^2  \approx \log \left(\frac{\mu_{M}}{\sigma_M} - \alpha \right).
		\end{equation} 
		This implies 
		\begin{footnotesize}
		\begin{align}\label{eq:a case2 log}
			\alpha &\approx -\sqrt{2} \sqrt{\log \frac{N-K}{K} -\log \left(\frac{\mu_{M}}{\sigma_M} - \alpha \right) - \log \sqrt{2\pi}\alpha^2 } \notag\\
			&\approx-\sqrt{2} \sqrt{\log \frac{(N-K)\sigma_M}{K\mu_M}}.
		\end{align}
		\end{footnotesize}
		Since in practice, neither $ N $ or $ M $ would be extremely large, $ \mu_{M}/\sigma_M $ is comparable or larger than $ \log N/K $, and therefore $ \alpha = -O\left(\log \frac{N \sigma_M}{K \mu_M}\right) $ from \eqref{eq:a case2 log}. 
		
		ii) In the second case $ \alpha \approx 0 $, which requires $ \frac{N}{K} \approx \frac{\mu_M}{\sigma_M}$ such that \eqref{eq:alphacase2} may hold. Applying Taylor expansion at $ \alpha = 0 $, we approximate \eqref{eq:alphacase2}
		with 
		\begin{equation}
			\frac{1}{\sqrt{2\pi}} + \frac{\alpha}{2} \approx \left(\frac{\mu_M}{\sigma_M} - \alpha\right) \frac{K}{N-K},
		\end{equation}
		implying 
		\begin{equation}
			\alpha \approx \frac{2(N-K)}{N+K} \left(\frac{\mu_M}{\sigma_M} \frac{K}{N-K} - \frac{1}{\sqrt{2\pi}}\right),
			\label{eq:alphascase3}
		\end{equation}
		or simply $ \alpha = O\left(\frac{\mu_M K}{\sigma_M N}\right)  $.

				iii) We finally consider the third case
		$ \alpha \gg 1 $. Substituting \eqref{eq:xerfc +} into \eqref{eq:taukktwrtalpha2} implies \vspace{-2ex}
		\begin{equation}
			\alpha \approx \frac{K\mu_M}{N\sigma_M} \gg 1,\vspace{-1ex}
			\label{eq:alphacase1}
		\end{equation}
		which requires that $ \frac{N}{K} \ll  \frac{\mu_M}{\sigma_M}$. 
		
		Summarizing the three cases above, we find that: i) When $ \frac{N}{K} \gg \frac{\mu_M}{\sigma_M}$, we have $ \alpha \ll -1 $ and \eqref{eq:a case2 log}; ii) When $ \frac{N}{K} \approx \frac{\mu_M}{\sigma_M}$, we have $ \alpha \approx 0$ and \eqref{eq:alphascase3}; iii) When $ \frac{N}{K} \ll \frac{\mu_M}{\sigma_M}$, we have \eqref{eq:alphacase1}.
		We note that these three cases are generally complete, representing three kinds of relationship between the relative sparsity ($ N/K $) and the number of available frequencies ($ \mu_{M}/ \sigma_M  \approx 2\sqrt{M} $).  
		With the obtained $ \alpha $, implying the miminizer $ \tau_{\ast} $, we next calculate the limit inferior $ x_b(\tau_{\ast}) $. \vspace{-2ex}

		\subsection{The infimum $ \inf_{\tau\ge 0} x_b(\tau) $}
		\label{subsec:app inf xb}
		Comparing $ x_b(\tau) $ in \eqref{eq:xbdetail} and $ \tau_{\ast} $ in \eqref{eq:taukkt}, we find the right hand side of \eqref{eq:taukkt} also appears in \eqref{eq:xbdetail}. We replace this term in \eqref{eq:xbdetail} by $ \tau_{\ast} $, so that \eqref{eq:xbdetail} becomes
		\begin{align}
			x_b(\tau_{\ast})  &= \tau_{\ast}^2 + \frac{N-K}{2K}\sigma_M^2 {\rm erfc}\left(\frac{\tau_{\ast}-\mu_M}{\sqrt{2}\sigma_M} \right) + (\mu_M-\tau_{\ast}) \tau_{\ast} \notag \\
			&= \mu_M \tau_{\ast} + \frac{N-K}{2K}\sigma_M^2 {\rm erfc}\left(\frac{\tau_{\ast}-\mu_M}{\sqrt{2}\sigma_M} \right) .
			\label{eq:xtauast}
		\end{align}
		Substituting $\tau_{\ast} =\mu_M - \alpha \sigma_M $, we rewrite \eqref{eq:xtauast} as
		\begin{small}
		\begin{align}
			x_b(\tau_{\ast})  &= \mu_M (\mu_M - \alpha \sigma_M) + \frac{N-K}{2K}\sigma_M^2 {\rm erfc}\left(- \frac{\alpha}{\sqrt{2}} \right). 
			\label{eq:xalpha}
		\end{align}
		\end{small}
		For the three cases considered below \eqref{eq:alphascase3}, we calculate the limit inferior $ x_b(\tau_{\ast}) $ with the obtained $ \alpha $, respectively. 
		
		1) In the first case when $ \frac{N}{K}  \gg \frac{\mu_M}{\sigma_M}$, we have $ \alpha \ll -1 $. Using Taylor expansion, we approximate \eqref{eq:xalpha} by
		\begin{equation}
			x_b(\tau_{\ast})  \approx \mu_M (\mu_M - \alpha \sigma_M) - \frac{N-K}{2K}\sigma_M^2 \frac{\sqrt{2}e^{-\frac{\alpha^2}{2} }}{{\sqrt {\pi }} \alpha},
		\end{equation}
		which can be further simplified by replacing the exponent term according to \eqref{eq:a case2}, given by\vspace{-2ex}
		\begin{align}\label{eq:xb case2}
			x_b(\tau_{\ast})  &\approx \mu_M (\mu_M - \alpha \sigma_M) - \frac{N-K}{2K}\sigma_M^2 \frac{\sqrt{2}e^{-\frac{\alpha^2}{2} }}{{\sqrt {\pi }} \alpha}\notag \\
			& = \mu_M (\mu_M - \alpha \sigma_M) + \alpha \sigma_{M}^2 \left(\alpha - \frac{\mu_M}{\sigma_M}\right) \notag \\
			& = (\mu_M - \alpha \sigma_M)^2.
		\end{align}
		
		Plugging \eqref{eq:muM2 app}, \eqref{eq:sigmaM app} and \eqref{eq:a case2 log} into \eqref{eq:xb case2}, we have\vspace{-2ex}
		\begin{align}
		    x_b(\tau_{\ast}) &\approx  \mu_M^2 - 2\alpha \mu_M \sigma_M \notag\\
		    &\approx 2M-\frac{1}{2} + \sqrt{2}\sqrt{(4M-1)\log \frac{(N-K)}{K\sqrt{4M-1}}}.
		\end{align}
		
		Thus, the final $N_b$ is 
		\begin{footnotesize}
		\begin{equation} \label{eq:nbdot}
		    N_b \approx N_{b1} = 2MK-\frac{K}{4}+\frac{\sqrt{2}K}{2}\sqrt{(4M-1)\log \frac{(N-K)}{K\sqrt{4M-1}}}.
		\end{equation}
		\end{footnotesize}

    		2) The second case corresponds to $ \frac{N}{K} \approx \frac{\mu_M}{\sigma_M}$, leading to $ \alpha \approx 0 $. Expanding the $ {\rm erfc}(\cdot) $ term at $ \alpha = 0 $, we approximate \eqref{eq:xalpha} by\vspace{-2ex}
    		\begin{small}
		\begin{align}
			x_b(\tau_{\ast})  &\approx \mu_M (\mu_M - \alpha \sigma_M) + \frac{N-K}{2K}\sigma_M^2 \left(1 - \sqrt{\frac{2}{\pi}} \alpha \right). \vspace{-3ex} 
			\label{eq:xalpha case3}
		\end{align}
		\end{small}
		
		Substituting \eqref{eq:alphascase3}, \eqref{eq:muM2 app} and \eqref{eq:sigmaM app} into \eqref{eq:xalpha case3} yields 
		\begin{align}
			&x_b(\tau_{\ast}) =   2M-\frac{3}{4}+\frac{N}{4K}-\frac{(4M-1)K}{N+K}\notag \\
			&\quad +\frac{(N-K)^2}{2K\pi (N+K)}-\frac{(\sqrt{2}+1)(N-K)}{2(N+K)} \sqrt{\frac{4M-1}{\pi}}.                                                                                                                                
		\end{align}
		
		Plugging the above result into \eqref{eq:Nb tau}, we have
		\begin{equation} \label{eq:nbddot}
			N_b \approx N_{b2} = KM + \frac{K}{2}x_b(\tau_{\ast}).
		\end{equation}
		
		3) In the third case when $ \frac{N}{K} \ll \frac{\mu_M}{\sigma_M}$, we substitute \eqref{eq:alphacase1} into \eqref{eq:xalpha}, and obtain\vspace{-2ex} 
		
		\begin{small}
		\begin{equation}
			\label{eq:xbcase1}
			x_b(\tau_{\ast}) = \frac{N-K}{N}\mu_{M}^{2}+\frac{N-K}{2K}\sigma_{M}^{2}{\rm erfc}\left(-\frac{K\mu_M}{\sqrt{2}N\sigma_M}\right).\vspace{-2ex} 
		\end{equation}
		\end{small}
		
		Plugging \eqref{eq:muM2 app} and \eqref{eq:sigmaM app} into \eqref{eq:xbcase1}, we have the approximation
		\begin{align}
			x_b(\tau_{\ast})&\approx\frac{N-K}{N}\left(2M-\frac{1}{2}\right)+\frac{N-K}{2K} \notag \\
			&=2M-1-\frac{K}{N}\left(2M-\frac{1}{2}\right)+\frac{N}{2K}.
			\label{eq:xbapcase1}
		\end{align}
		
		With \eqref{eq:xbapcase1}, we rewrite \eqref{eq:Nb tau} approximately as
		\begin{small}
			\begin{align}
				\label{eq:nb1}
				N_b &\approx N_{b3} = KM + \frac{K}{2}\cdot \left(2M-1-\frac{K}{N}\left(2M-\frac{1}{2}\right)+\frac{N}{2K}\right) \notag \\
				& =  \left(2M-\frac{1}{2}\right)K-\frac{(4M-1)K^2}{4N}+\frac{N}{4} \notag\\
				&=MN-\frac{(4M-1)(N-K)^2}{4N}.\vspace{-4ex}
			\end{align}
			\end{small}
		
		Summarizing the three cases above, we have that i) $\frac{N}{K} \ll \frac{\mu_M}{\sigma_M}$, ii) $\frac{N}{K} \approx \frac{\mu_M}{\sigma_M}$, or iii) $\frac{N}{K} \gg \frac{\mu_M}{\sigma_M}$, $N_b$ is approximated by \eqref{eq:nbdot}, \eqref{eq:nbddot} or \eqref{eq:nb1}, respectively, completing the proof of Proposition~\ref{pro:appprx nb}.\vspace{-2ex}

		\section{Proof of Proposition \ref{pro:xs app}} 
		\label{proof:xs}
		For notation purposes, we use $x_s(\tau)$ to represent the term inside the limit inferior operation of \eqref{equa:solvesSR} as
		\begin{align}
			x_s&(\tau):=\tau^2 + \frac{N-K}{K}\int_{\tau}^\infty\!\!\!(u\!-\!\tau)^2\phi_{2}(u)\mathrm{d}u \label{eq:xs 0} \\
			&=\tau^2 + \frac{N-K}{K} \int_{\tau}^\infty \!\!\! (u\!-\!\tau)^2 u e^{-u^2/2}\mathrm{d}u \notag\\
			&\stackrel{(a)}{=}\tau^2+\frac{N-K}{K}\left[2e^{-\tau^2/2}-\sqrt{2\pi}\tau{\rm erfc}(\tau/\sqrt{2})\right]
			,		\label{eq:xs}
		\end{align}
		where (a) comes from \eqref{eq:normal int4}, \eqref{eq:normal int1} and \eqref{eq:normal int2}. Then, $ N_s $ in \eqref{equa:solvesSR} is given by \vspace{-2ex}
		\begin{equation}
			N_s = KM + \frac{KM}{2} \inf_{\tau\ge 0} x_s(\tau). \vspace{-1ex}
			\label{eq:Ns}
		\end{equation}
		
		Taking partial derivatives over both sides of \eqref{eq:xs 0} and letting $ \frac{\partial x_s(\tau)}{\partial \tau} =0 $, we find the minimizer that leads to the limit inferior of $ x_s(\tau) $\vspace{-2ex}
		\begin{align}
			\tau_{\star} &= \frac{N-K}{K}\int_{\tau_{\star}}^\infty\!\!\!(u\!-\!\tau_{\star})\phi_{2}(u)\mathrm{d}u  \notag \\
			&=\frac{N-K}{K}\int_{\tau_{\star}}^\infty\!\!\!(u\!-\!\tau_{\star})u e^{-u^2/2}\mathrm{d}u\notag\\ 
			& \stackrel{(a)}{=} \frac{N-K}{K}\cdot \sqrt{\frac{\pi}{2}}{\rm erfc}\left(\tau_{\star}/\sqrt{2}\right),\vspace{-2ex}
			\label{eq:kkts}
		\end{align}
		where (a) holds according to \eqref{eq:normal int1} and \eqref{eq:normal int2}. 
		Similarly to the technique used in \ref{proof:xb}, we first approximately solve \eqref{eq:kkts}, and then calculate $ x_s(\tau_{\star}) $. 
		
		The value $ \tau_{\star} $ relies on $ N $ and $K$. Since $ N \ge K $ (otherwise the unique recovery of the $ K $ extended targets is not possible), we consider two cases i) $   \frac{N}{K} \approx 1$ and ii) $  \frac{N}{K} \gg 1$, representing the less sparse and relatively sparse cases, respectively. 
		
		i)  In the first case, $   \frac{N}{K} \approx 1$, it is deduced from \eqref{eq:kkts} that $ \tau_{\star}$ takes values around 0. 
		Approximating $ {\rm erfc}\left(\tau / \sqrt{2}\right)$ at $ \tau = 0 $ with first order Taylor expansion, 
		we rewrite \eqref{eq:kkts} as \vspace{-1ex}
		\begin{equation}
		    \tau_{\star} \approx \frac{N-K}{K}\cdot (\sqrt{{\pi}/{2}}-\tau_{\star}),\vspace{-1ex}
		\end{equation} 
		which implies 
		\begin{equation}\vspace{-1ex}
		    \tau_{\star} \approx \frac{N-K}{N}\cdot \sqrt{{\pi}/{2}}.
			\label{eq:tau s dense}
		\end{equation}	\vspace{-2ex}

		ii) In the second case, when $  \frac{N}{K} \gg 1$, we have that $\tau_{\star}$ is also sufficiently large $ \tau_{\star} \gg 1$. Therefore, we expand the $ {\rm erfc}(\cdot) $ function at $ +\infty $, and rewrite \eqref{eq:kkts} as 
		\begin{align}
			\label{eq:tau s sparse}
			\tau_{\star} &\approx \frac{N-K}{K}\cdot \sqrt{\frac{\pi}{2}} \cdot \sqrt{\frac{2}{\pi}} e^{-{\tau_{\star}^2}/2} (\frac{1}{\tau_{\star}}-\frac{1}{\tau_{\star}^3}) \notag\\
			&\approx \frac{N-K}{K}e^{-{\tau_{\star}^2}/2}\frac{\tau_{\star}}{\tau_{\star}^2+1}
			, 
		\end{align}
		implying 
		\begin{equation}
			\label{eq:tau s sparse2}
			 \log (\tau_{\star}^2+1) \approx \log \frac{N-K}{K} -\frac{\tau_{\star}^2}{2}.
		\end{equation}
		This yields $ \tau_{\star} = O \left(\sqrt{\log (N/K)}\right) $.
		
		We then calculate $ x_s(\tau_{\star}) $ with the substitution of $ \tau_{\star} $. 
		Note that the erfc term in \eqref{eq:xs} can be replaced by linear term according to \eqref{eq:kkts}, which simplifies \eqref{eq:xs} into\vspace{-1ex}
		\begin{equation}
			\label{eq:xst}
			x_s(\tau_{\star})=-{\tau_{\star}}^2+\frac{N-K}{K}\cdot 2e^{-\tau_{\star}^2/2}
			.\vspace{-1.5ex}
		\end{equation}
		We analyze the results in both cases i) $  \frac{N}{K} \gg 1$ and ii) $   \frac{N}{K} \approx 1$, respectively. 
		
		1) when $\frac{N}{K}\gg 1$, we have that $\tau_{\star}$ is sufficiently large. 
		Replacing the exponent term in \eqref{eq:xst} with a quadratic term according to \eqref{eq:tau s sparse}, we simplify \eqref{eq:xst} into \vspace{-1.5ex}
		\begin{equation}
			\label{eq:xstapp}
			x_s(\tau_{\star})\approx \tau_{\star}^2+2.\vspace{-1.5ex}
		\end{equation}
		Considering \eqref{eq:tau s sparse2}, we substitute \eqref{eq:xstapp} into \eqref{eq:Ns}, yielding\vspace{-1ex}
		\begin{equation}
			\label{eq:nsg1}
			N_s \approx N_{s1} = 2MK + \frac{MK\tau_{\star}^2}{2},\vspace{-2ex}
		\end{equation}
		where $\tau_{\star}^2$ can be calculated from \eqref{eq:tau s sparse2}. 
		
		2) 
		When $   \frac{N}{K} \approx 1$, we have $ \tau_{\star} \approx 0 $, thus we approximate $e^{-\tau_{\star}^2/2}$ at $ \tau_{\star} = 0 $, i.e., $ \left. e^{-\tau_{\star}^2/2} \right|_{\tau_{\star}=0}\approx 1-\frac{\tau_{\star}^2}{2} $, and rewrite \eqref{eq:xst} as\vspace{-1ex} 
		\begin{equation}
		x_s(\tau_{\star}) \approx -\frac{N\tau_{\star}^2}{K}+\frac{2(N-K)}{K}.
			\label{eq:xs app dense}
		\end{equation}
		Substituting \eqref{eq:tau s dense} into \eqref{eq:xs app dense} yields\vspace{-1ex} 
		\begin{equation}
			\label{eq:xst2}
				x_s(\tau_{\star}) \approx -\frac{\pi(N-K)^2}{2NK}+\frac{2(N-K)}{K}.
		\end{equation}
		Together with \eqref{eq:Ns}, this implies \vspace{-1ex} 
		\begin{equation}
			\label{eq:ns1}
			N_s \approx N_{s2} = MN-\frac{\pi M(N-K)^2}{4N}.\vspace{-1ex}
		\end{equation}
		
		Combining the two cases above, we have approximations of $N_s$ given by \eqref{eq:nsg1} or  \eqref{eq:ns1}, respectively when i) $\frac{N}{K} \ll 1$ or ii) $\frac{N}{K} \approx 1$ holds, completing the proof of Proposition~\ref{pro:xs app}.

	\end{appendices}
	
	\bibliographystyle{ieeetr}
	\bibliography{phase_transition}
	
\end{document}